\documentclass[prd,12pt]{revtex4}

\usepackage{amssymb,amsthm,psfig}

\newcommand{\cD}{{\mathcal D}}

\newcommand{\beq}{\begin{equation}}
\newcommand{\eeq}{\end{equation}}

\newcommand{\ber}{\begin{eqnarray}}
\newcommand{\eer}{\end{eqnarray}}

\newcommand{\su}{{\mathfrak su}}
\newcommand{\mb}{{\bar{m}}}

\newcommand{\tlX}{\tilde{X}}

\newcommand{\tr}{{\rm Tr}}

\def\nm{c}

\begin{document}

\title{Non-Metric Gravity II: Spherically Symmetric Solution, \\
Missing Mass and Redshifts of Quasars}

\date{May 14, 2007}

\author{Kirill Krasnov\,$^a$ and Yuri Shtanov\,$^b$}
\affiliation{$^a$School of Mathematical Sciences, University of Nottingham,
Nottingham, NG7 2RD, UK \& Perimeter Institute for Theoretical Physics,
Waterloo, N2L 2Y5, Canada. \\ $^b$Bogolyubov Institute for Theoretical Physics,
Kiev 03680, Ukraine.}

\begin{abstract}
We continue the study of the non-metric theory of gravity introduced in
hep-th/0611182 and gr-qc/0703002 and obtain its general spherically symmetric
vacuum solution.  It respects the analog of the Birkhoff theorem, i.e., the
vacuum spherically symmetric solution is necessarily static. As in general
relativity, the spherically symmetric solution is seen to describe a black
hole. The exterior geometry is essentially the same as in the Schwarzschild
case, with power-law corrections to the Newtonian potential. The behavior
inside the black-hole region is different from the Schwarzschild case in that
the usual spacetime singularity gets replaced by a singular surface of a new
type, where all basic fields of the theory remain finite but metric ceases to
exist. The theory does not admit arbitrarily small black holes: for small
objects, the curvature on the would-be horizon is so strong that non-metric
modifications prevent the horizon from being formed.

The theory allows for modifications of gravity of very interesting nature. We
discuss three physical effects, namely, (i)~correction to Newton's law in the
neighborhood of the source, (ii)~renormalization of effective gravitational and
cosmological constants at large distances from the source, and (iii)~additional
redshift factor between spatial regions of different curvature.  The first two
effects can be responsible, respectively, for the observed anomaly in the
acceleration of the Pioneer spacecraft and for the alleged missing mass in
spiral galaxies and other astrophysical objects. The third effect can be used
to propose a non-cosmological explanation of high redshifts of quasars and
gamma-ray bursts.
\end{abstract}

\maketitle

\section{Introduction}
\label{sec:intro}

This is the second paper in the series devoted to the detailed study of the
non-metric theory of gravity introduced in
\cite{Krasnov:2006du,Krasnov:2007uu}. As in the case of general relativity
(GR), one of the first applications of any theory of gravity must be the
description of the geometry of a strongly gravitating object --- a black hole.
This is the basic aim of the present paper, in which we obtain and analyze the
spherically symmetric vacuum solution. The considerations of this work will
also serve as an illustration to the rather abstract formalism of
\cite{Krasnov:2007uu}.

The theory of gravity under investigation is defined by the action
\ber\label{action}
S[B,A,\Psi]= \frac{1}{8\pi G} \int_M B^i \wedge F^i + \frac{1}{2} \left(
\Psi^{ij} + \phi \, \delta^{ij} \right) B^i \wedge B^j \, , \\ \nonumber
\phi=\phi \left[ \tr \left(\Psi^2 \right), \tr \left(\Psi^3 \right) \right] \,
.
\eer
Here, $B^i$ is a (complex) $\su(2)$ Lie-algebra valued two-form (the indices
$i,j = 1,2,3$ belong to the $\su(2)$ Lie algebra), $F^i=dA^i + \frac12 [A,A]^i$
is the curvature of the $\su(2)$ Lie-algebra valued connection $A^i$,
$\Psi^{ij}$ is a traceless symmetric ``Lagrange multiplier'' field, and $G$ is
the Newton's constant. The function $\phi \left[ \tr \left( \Psi^2 \right), \tr
\left( \Psi^3 \right) \right]$ is a function of two scalar invariants that can
be constructed from $\Psi^{ij}$ and is responsible for the departure of the
theory under consideration from GR. The constant part of the function $\phi$ is
(a multiple of) the usual cosmological constant $\Lambda$.

The theory with action (\ref{action}) modifies the Pleba{\'n}ski self-dual
formulation of general relativity \cite{Plebanski:1977}, in which $\phi \equiv
\phi_0 = - \Lambda / 3$.  In the Pleba{\'n}ski theory, the field $B^i$,
according to the equations of motion, can be canonically decomposed into
products of tetrad basis one-forms, which then define a unique distinguished
metric satisfying the Einstein equation.  With the appearance of a nontrivial
function $\phi(\Psi)$ in (\ref{action}), this property is no longer valid: the
arising metric is defined only up to a conformal factor, and it does not, in
general, satisfy the vacuum Einstein equations.

The nature of this modification of gravity is quite special and deserves a few
words. In the numerous existing schemes of modified gravity, one usually deals
with a metric theory and modifies the Hilbert--Einstein action by introducing
extra degrees of freedom --- either by increasing the number of derivatives
(higher-derivative gravity), or by introducing extra fields
(scalar-vector-tensor theories), or by considering extra dimensions
(braneworlds).  Nothing of the listed takes place in our generalization: the
theory remains four-dimensional, there are no extra fields, the number of
derivatives is not increased, and, as in GR, there are still just two
propagating degrees of freedom, as can be seen from the canonical analysis of
the theory; see, e.g., \cite{Bengtsson:2007zx}. This is achieved by first
recasting the theory in a form which does not contain any metric
\cite{Plebanski:1977}, and then modifying the theory in this non-metric form.
It is thus important to stress that the term ``non-metric'' is used here not in
the sense that some additional degrees of freedom are present along with the
usual metric, but in the sense that metric does not even appear in the
formulation of the theory.

The difference between our modified gravity and a variety of other approaches
can also be seen from the fact that our theory, while modifying the spherically
symmetric solution, does not modify the formal general-relativistic
cosmological equations. Indeed, due to the high symmetry, the field $\Psi$,
which is a close analog of the Weyl curvature spinor in our theory, is
identically zero for cosmological solutions, which then coincide with those of
general relativity with a cosmological constant determined by $\phi(0)$.
However, any departures from homogeneity and isotropy will be essential, so
that the evolution of perturbations in our theory will, in principle, be
different from that in the concordance LCDM cosmology.  This issue will be
studied separately.

A specific choice of the function $\phi$ uniquely fixes a theory from the class
(\ref{action}). One way of fixing the form of this function is to regard it as
an effective quantum contribution to the original classical action. In this
respect, it was argued in \cite{Krasnov:2006du} that the class of theories
(\ref{action}) is closed under the renormalization-group flow. The conjecture
of asymptotic safety applied to the case at hand then asserts the existence of
a non-trivial ultra-violet fixed point of the renormalization-group flow
described by a certain function $\phi^*(\Psi)$. It would then make sense to
choose this fixed-point function $\phi^*(\Psi)$ in (\ref{action}) because the
{\em quantum\/} field theory so defined would have extremely appealing
properties: the action would not get perturbatively renormalized, describing an
essentially finite quantum theory of gravity. This gives a possible scenario
for fixing $\phi(\Psi)$ from the theoretical side.

In a pure quantum theory of gravity, the Planck scale is the only scale that
can enter the renormalized action. However, once the theory is coupled to
matter, the corresponding quantum loops in Feynman diagrams will also affect
the function $\phi$. Thus, it should be expected that the ultra-violet
fixed-point function $\phi^*$ will contain more than one physical scale.
Qualitative features of the function $\phi^*$ can be anticipated from the fact
that, as we shall see in this paper, a theory with a non-constant function
$\phi(\Psi)$ in many respects behaves like a theory with physical parameters
depending on the scale. Indeed, as we shall see, the function $\phi(\Psi)$
itself will receive an interpretation of a curvature-dependent cosmological
``constant''. Another effect is that passing between spatial regions of
different curvature generally introduces a renormalization in the strength of
the gravitational interaction. All this is strongly suggestive of the
renormalization-group flow phenomena, where one has to identify $\Psi$ (having
the dimension of the curvature) with the energy scale squared. The crucial
difference between our scheme and the renormalization-group flow familiar from
the framework of effective field theory is that similar effects arise here in a
diffeomorphism-invariant context. This renormalization-group interpretation
suggests that the function $\phi^*(\Psi)$ must look as a sequence of plateaus,
with crossover regions between plateaus corresponding to the scales where new
physics (new degrees of freedom) come into play. This discussion motivates some
assumptions we make about the form of the function $\phi (\Psi)$ in the section
where we discuss possible long distance modifications of gravity.

Of special importance is the question of coupling our theory of gravity
(\ref{action}) to other fields. As was discussed in \cite{Krasnov:2007uu},
coupling to Yang-Mills fields (or electromagnetic field) is seamless, and the
action described in \cite{CDJ} extends to the non-metric situation without any
problem. This action tells us how massless particles such as photons interact
with gravity and predicts their motion in a non-metric background. It is found
that, in the approximation of geometric optics, photons move along null
geodesics of the metric determined by the field $B^i$. The ambiguity in the
choice of the conformal factor discussed in detail in \cite{Krasnov:2007uu}
does not affect this conclusion because the paths of null geodesics (but, of
course, not the affine parameter along them) are independent of this choice. An
important issue that has not yet been addressed in the framework of non-metric
gravity and that prevents us from a complete analysis of the physical
predictions of theory (\ref{action}) concerns coupling to (massive) matter
degrees of freedom. The action for a massive field of spin $1/2$ proposed in
\cite{CDJ} for the Pleba{\'n}ski formulation of general relativity turns out to
be incompatible with the non-metric character of the theory under
investigation, hence, calls for revision.

Quantitative predictions of our theory, e.g., concerning the motion of stars
and gas in galaxies, will depend on the details of the coupling of massive
matter to the basic gravitational degrees of freedom.  However, its certain
general features can already be described in the absence of these details. For
instance, in the domains of ``metricity,'' in which the function $\phi$ is
almost constant (where its dimensionless derivatives $|\partial \phi / \partial
\Psi| \ll 1$), the present theory of gravity behaves very closely to general
relativity, and it is quite reasonable to expect that matter will also behave
accordingly, moving relativistically in the background of the arising metric.
Assuming that several such domains of ``metricity'' exist at different scales
of curvatures, we find that the effective gravitational mass of a central body
is different in the corresponding spatial regions. From this simple observation
one can conclude that the effective gravitational mass of a body continuously
depends on the distance to this body even in the case of general function $\phi
(\Psi)$  --- the effect of scale-dependence of the gravitational coupling which
was mentioned above. One can use this property to account for the phenomenon of
missing mass observed in gravitating objects such as spiral and elliptical
galaxies.  Another interesting general prediction of the present theory is the
appearance of an additional redshift factor between regions of different
space-time curvature. As we point out in this paper, this effect can be used to
account for the observed high redshifts of quasars and gamma-ray bursts.
However, practical use of these features of the new theory to explain physical
phenomena requires the specific knowledge of the underlying function $\phi
(\Psi)$ together with the analysis of the physical content and interpretation
of the theory.  This will be the subject of the future work.

The organization of the paper is as follows. To obtain a spherically symmetric
solution, we first obtain an ansatz for the symmetric field $B^i$. This is done
in Sec.~\ref{sec:ansatz}, in which we also analyze consequences of the modified
``metricity'' equations.  The field equations are obtained in
Sec.~\ref{sec:eqs}. Solutions are analyzed and interpreted in
Sec.~\ref{sec:solution}. Possible modifications of gravity are discussed in
Sec.~\ref{sec:mod}, and possible physical effects in Sec.~\ref{sec:effects}.
Our results are summarized in Sec.~\ref{sec:sum}.  In the appendix, we give a
detailed proof of the static property of the spherically symmetric solution in
the theory under consideration and discuss in more generality some details of
the large-distance modifications of gravity.

\section{Spherically symmetric ansatz and metricity equations}
\label{sec:ansatz}

\subsection{Spherically symmetric Lie-algebra-valued two-form}

The most general spherically symmetric $\su(2)$ Lie-algebra-valued two-form can
be obtained from the condition that an ${\rm SO}(3)$ rotation corresponds to a
gauge transformation.  There exists standard technique in the literature
allowing one to obtain the relevant expression; see, e.g.,
\cite{Brodbeck:1996ma}. One gets:
\beq \label{B-symm}
\begin{array}{rcl}
B \equiv \sum_i B^i \tau^i  &=& \left(\phi_1 dt\wedge d\theta - \chi_1
\sin\theta\, dr\wedge d\phi \right) \tau^1  \\ &+& \left( \phi_2 \sin\theta\,
dt\wedge d\phi + \chi_2 dr\wedge d\theta \right) \tau^2 \\
&+& \left( \phi_3 \sin\theta\, d\theta\wedge d\phi + \chi_3 dt\wedge dr \right)
\tau^3 \, .
\end{array}
\eeq
Here, $(t, r, \theta, \phi)$ is the standard set of spherical coordinates, and
$\phi_i$, $\chi_i$, $i=1,2,3$, are functions of $t$ and $r$ only. The symbols
$\tau^i$ denote the $\su(2)$ generators $\tau^i=-(i/2)\sigma^i$, where
$\sigma^i$ are the Pauli matrices. As it turns out, it is much more convenient
to work not with the adjoint, but with the fundamental representation of
$\su(2)$. This amounts to working in the spinor formalism. We used spinor
formalism rather heavily in \cite{Krasnov:2007uu}, and will continue to do so
in this paper. To pass from the ${\rm SO}(3)$ form of the fields to their
spinor representation one has to replace every lower-case Latin index
$i,j,\ldots = 1,2,3$ by a symmetric pair of unprimed spinor indices.
Equivalently, every $\su(2)$ Lie-algebra-valued field gets replaced by a
$2\times 2$ matrix-valued field. Thus, it will be convenient to rewrite the
above ansatz for $B$ in terms of the matrices
\beq
\tlX_- = \left(
\begin{array}{rr} 0 & 0 \\ -1 & 0 \end{array} \right), \qquad \tlX_+ = \left(
\begin{array}{rr} 0 & 1 \\ 0 & 0 \end{array} \right), \qquad \tlX = \frac{1}{2}
\left( \begin{array}{rr} 1 & 0 \\ 0 & -1
\end{array} \right) \, ,
\eeq
which are related to $\tau^i$ as $\tau^1 = (1/2i) ( \tlX_+ - \tlX_- )$, $\tau^2
= (1/2) ( \tlX_+ + \tlX_- )$, $\tau^3 = -i\tlX$. We have:
\beq \label{B-1}
\begin{array}{rcl}
B &\equiv& \tlX_- B_- + \tlX_+ B_+ + \tlX B_0  \medskip \\
&=& \displaystyle \tlX_-\left[ \left( - \frac{1}{2i} \phi_1 dt + \frac{1}{2}
\chi_2 dr\right) \wedge d\theta + \left( \frac{1}{2i} \chi_1 dr + \frac{1}{2}
\phi_2 dt \right) \sin\theta \wedge d\phi \right] \medskip \\
&+& \displaystyle \tlX_+\left[ \left( \frac{1}{2i} \phi_1 dt + \frac{1}{2}
\chi_2 dr\right) \wedge d\theta + \left( - \frac{1}{2i} \chi_1 dr + \frac{1}{2}
\phi_2 dt \right) \sin\theta \wedge d\phi \right] \medskip \\
&-& i\tlX \left( \phi_3 \sin\theta\,  d\theta \wedge d\phi + \chi_3 dt\wedge dr
\right)\, .
\end{array}
\eeq

It is not hard to show, and this fact was used heavily in \cite{Krasnov:2007uu}, that,
by a convenient choice of the spinor basis, the spinor counterpart of the
quantity $\Psi$ can be put into the form
\beq \label{psi-gen}
\Psi = \alpha \left( \tlX_- \otimes \tlX_- + \tlX_+ \otimes \tlX_+ \right) +
\beta \left( \tlX_+ \otimes \tlX_- + \tlX_-\otimes \tlX_+ + 4 \tlX\otimes \tlX
\right) \, .
\eeq
The functions $\alpha$ and $\beta$ are related to the invariant characteristics
of $\Psi\,$:
\beq \label{ab}
\tr \left( \Psi^2 \right) = 2 \alpha^2 + 6 \beta^2 \, , \qquad \tr \left(
\Psi^3 \right) = 6 \beta \left( \alpha^2 - \beta^2 \right) \, .
\eeq
In view of (\ref{ab}), one can regard $\phi \left[ {\rm Tr} \left( \Psi^2
\right),{\rm Tr} \left(\Psi^3 \right) \right]$ as a function of $\alpha$ and
$\beta$.

In the spherically symmetric case, the field $\Psi^{ij}$ has the form
\beq
\Psi^{ij} = \psi (r) \left( x^i x^j - \frac13 \delta^{ij} r^2 \right) \, ,
\eeq
where $x^i$, $i = 1,2,3$, are the natural Euclidean coordinates realizing the
group of rotations, and $r^2 = \sum_i \left(x^i\right)^2$.  This condition
implies $\alpha = 0$ in (\ref{psi-gen}), and the field $\Psi$ is parameterized
by a single function $\beta\,$:
\beq\label{psi}
\Psi = \beta \left( \tlX_+ \otimes \tlX_- + \tlX_-\otimes \tlX_+ + 4
\tlX\otimes \tlX \right) \, .
\eeq
In other words, the Lagrange multiplier field $\Psi$ must be algebraically
special, of type $D$. This is, of course, exactly the property of $\Psi$ in a
spherically symmetric solution of the usual GR.  This property remains
unchanged in the non-metric theory of gravity under consideration.

\subsection{``Metricity'' equations}

With the above ansatz for the $B$ field, it is easy to compute the quantity
$B^i\wedge B^j$ that appears in the metricity equations --- the equations
stemming from (\ref{action}) as the field $\Psi$ is varied. These equations are
discussed in \cite{Krasnov:2007uu} in great length, and we will not repeat that
discussion here. We just note that, as it is easy to check, the quantity $B^i
\wedge B^j$, with $B^i$ given by (\ref{B-symm}) above, is diagonal as a
$3\times 3$ matrix. This implies that the matrix
$\Phi:=\partial\phi/\partial\Psi$, which is the traceless part of $B^i\wedge
B^j$ in view of the metricity equation, is also diagonal. Hence, $\phi_\alpha :
= \partial \phi / \partial \alpha$ must be identically zero in view of equation
(26) of \cite{Krasnov:2007uu}. The property $\alpha = 0$ ensures this
condition, in particular, if the function $\phi$ is a regular function of its
arguments (\ref{ab}) in the neighborhood of zero.

Let us now write the metricity equations specialized to the type $D$ at hand.
They are easy to obtain from equations (29) of \cite{Krasnov:2007uu} by setting
$\phi_\alpha = 0$. We have
\beq \label{metricity}
\begin{array}{l}
B_+ \wedge B_+ = B_- \wedge B_- = B_+ \wedge B_0 = B_- \wedge B_0 = 0\, , \medskip \\
\displaystyle 2 B_+ \wedge B_- + B_0 \wedge B_0 = - 2 \phi_\beta \left( B_+
\wedge B_- - \frac{1}{4} B_0 \wedge B_0 \right) \, ,
\end{array}
\eeq
where $\phi_\beta : = \partial \phi / \partial \beta$. Let us now see what this
implies about our ansatz (\ref{B-1}). The last two equations in the first line
of (\ref{metricity}) are automatically satisfied, while the first two equations
imply $\phi_1\chi_1=\phi_2\chi_2$. This allows us to write the $\tlX_\pm$
components in (\ref{B-1}) as
\beq
B_- = \nm\, m \wedge l \, , \qquad B_+ = \nm\, n \wedge \mb \, ,
\eeq
where
\beq
\label{forms}
\begin{array}{ll}
m := \displaystyle \xi \left(- \frac{\phi_1}{\phi_2} d\theta + i \sin\theta\,
d\phi \right) \, , \quad &l := \displaystyle \eta \left( -\frac{1}{2i} \phi_2
dt + \frac{1}{2}\chi_1 dr \right) \, , \medskip \\ \mb := \displaystyle \xi
\left(- \frac{\phi_1}{\phi_2} d\theta - i \sin\theta\,  d\phi \right) \, ,
\quad &n := \displaystyle \eta \left( -\frac{1}{2i} \phi_2 dt -
\frac{1}{2}\chi_1 dr \right) \, .
\end{array}
\eeq
In these expressions, $\xi$ and $\eta$ are arbitrary functions of $t$ and $r$,
and $\nm = \xi \eta$. To fix these functions, we equate the component $B_0$ in
(\ref{B-1}) to the third canonical two-form:
\beq
-i \left( \phi_3 \sin\theta\, d\theta \wedge d\phi + \chi_3 dt\wedge dr \right)
= l\wedge n - m\wedge \mb \, ,
\eeq
which gives $\xi^2$ and $\eta^2$ in terms of $\phi_i$ and $\chi_i$\,:
\beq \label{xieta}
\xi^2 = {\phi_2 \phi_3 \over 2 \phi_1} \, , \qquad \eta^2 = {2 \chi_3 \over
\phi_2 \chi_1}\, .
\eeq
Eventually, the expression for $B$ takes the form
\beq\label{b}
B = \tlX_- \nm\, m\wedge l + \tlX_+ \nm\, n\wedge \mb + \tlX \left( l\wedge n -
m\wedge \mb \right) \, .
\eeq
The last metricity equation [the second line of (\ref{metricity})] then relates
the function $\nm$ to the function $\beta$ in (\ref{psi}) through a derivative
of $\phi\,$:
\beq\label{c}
\nm^2 = \frac{1 - \phi_\beta/2}{1 + \phi_\beta} \, .
\eeq

Now we are going to simplify the expressions for one-forms (\ref{forms}).
First, we can choose $\xi$ as a new radial coordinate. After this, introducing
new functions $f$, $g$, and $h$, one can write the one-forms $l$, $n$, $m$, and
$\mb$ as
\beq\label{frame1}
l=\frac{1}{\sqrt{2}} \left( fdt - gdr \right), \quad n = \frac{1}{\sqrt{2}}
\left( fdt + gdr \right), \quad m,\, \mb = \frac{r}{\sqrt{2}} \left( h d\theta
\pm i \sin\theta\,  d\phi \right).
\eeq

By solving the system of field equations, one can prove that the function $h$
is just a constant, and that the sought functions $f$, $g$, and $\beta$ are
independent of time. We demonstrate this property in the appendix.  Then, by
rescaling the angle $\phi$ and the radial coordinate $r$, we can set $h$ to be
identically equal to unity, after which the canonical set of one-forms is
expressed as
\beq\label{frame}
l=\frac{1}{\sqrt{2}} \left( fdt - gdr \right), \quad n = \frac{1}{\sqrt{2}}
\left( fdt + gdr \right), \quad m,\, \mb = \frac{r}{\sqrt{2}} \left( d\theta
\pm i \sin\theta\,  d\phi \right),
\eeq
and the metric $ds^2 = 2l\otimes n - 2m\otimes \mb$ defined by tetrad
(\ref{frame}) assumes the standard form
\beq\label{metric}
ds^2  = f^2 dt^2 - g^2 dr^2 - r^2 \left( d\theta^2 + \sin^2\theta\, d\phi^2
\right)
\eeq
of the spherically symmetric problem of general relativity. In this form, we
assume all space-time coordinates to be real.  Thus, the only novelty as
compared to the GR case is the presence of the function $\nm$ in the first two
terms of (\ref{b}).  The function $\nm$ is related to $\beta$ through the
derivative $\phi_\beta(\beta)$ of $\phi (\beta)$ via (\ref{c}). Note that, if
$\phi_\beta = 0$, we have $\nm = 1$, and the two-form $B$ reduces to that of
the GR (with a cosmological constant determined by the constant value of
$\phi$).

\section{Field equations}
\label{sec:eqs}

The theory under consideration respects the analog of Birkhoff's theorem. The
proof of the static property of the metric is given in the appendix; its
asymptotic flatness is demonstrated below. In the main text, we just assume the
tetrad forms to be given by expression (\ref{frame}), and the $\Psi$ field by
expression (\ref{psi}), with $f$, $g$, and $\beta$ being functions of the
radial coordinate $r$ only, and we will be looking for solutions of these
functions.

\subsection{Structural equations}

The Cartan structural equations (called compatibility equations in
\cite{Krasnov:2007uu}) give an algebraic relation between the
connection form $A$ and the two-form $B$ and its exterior derivative, which
allows one to solve these equations with respect to $A$.  As the first step
towards this solution, we obtain manageable expressions for the quantities
$dB_\pm$ and $dB_0$, where
\beq \label{bs}
B_- = \nm\, m\wedge l \, , \qquad B_+ = \nm\, n\wedge \mb, \qquad B_0 = l\wedge
n - m\wedge \mb \, ,
\eeq
and the one-forms $l$, $n$, $m$, $\mb$ are given by (\ref{frame}). This is an
easy exercise in differentiation, similar to what one does in obtaining the
Ricci rotation coefficients for metric (\ref{metric}). After simple
calculation, we obtain
\beq \label{bs-2}
\begin{array}{l}
\displaystyle dB_- = {1 \over \sqrt2 r} \left[ {(r \nm f)' \over gf}\,
l \wedge n \wedge m - \nm\, \cot\theta\, l \wedge m \wedge \mb \right] \, , \medskip \\
\displaystyle dB_+ = - {1 \over \sqrt2 r} \left[ {(r \nm f)' \over gf}\,
l \wedge n \wedge \mb + \nm\, \cot\theta\, n \wedge m \wedge \mb \right] \, , \medskip \\
\displaystyle dB_0 = \frac{\sqrt2}{rg}\, (l - n) \wedge m \wedge \mb \, ,
\end{array}
\eeq
where the prime denotes differentiation with respect to $r$.

Another way of expressing the results for $dB_\pm$, $dB_0$ is to project the
arising three-forms onto the basis of dual one-forms. This is conveniently done
by introducing, for an arbitrary four-form $C$, the scalar quantity $C_v$
defined as
\beq\label{comp}
C = C_v\,  l \wedge n \wedge m \wedge \mb \, .
\eeq
We get
\beq \label{db-comp}
\begin{array}{l}
\displaystyle \left( dB_- \wedge \mb \right)_v = \left( dB_+ \wedge m \right)_v
= \frac{(r \nm f)'}{\sqrt{2}\, r g f} \, , \medskip \\
\displaystyle - \left( dB_- \wedge n \right)_v = \left( dB_+ \wedge l \right)_v
= \frac{\nm}{\sqrt2\, r} \cot\theta \, ,
\medskip \\
\displaystyle \left( dB_0 \wedge n \right)_v = \left( dB_0 \wedge l \right)_v =
\frac{\sqrt2}{rg} \, ,
\end{array}
\eeq
with all other components being zero.

The Cartan structural equations that determine the
components $A_\pm$, $A_0$ of the connection have the form
\beq\label{eqs-conn}
\begin{array}{l}
C_- = A_- \wedge B_0 - A_0 \wedge B_- \, , \\
C_+ = A_0 \wedge B_+ - A_+ \wedge B_0 \, , \\
C_0 = A_-\wedge B_+ - A_+ \wedge B_- \, ,
\end{array}
\eeq
with
\beq \label{cs}
C_\pm = - dB_\pm \, , \qquad C_0 = - \frac12 dB_0 \, ;
\eeq
see \cite{Krasnov:2007uu} for derivation. These equations are obtained from the equations
in \cite{Krasnov:2007uu} by specializing to the case of a ``constant'' basis in the
space of ``internal'' spinors, see \cite{Krasnov:2007uu} for a description of the
distinction between ``internal'' and ``spacetime'' spinors. The basis in which the
Lagrange multiplier has the form (\ref{psi}) is precisely such a basis.

These equations are solved by first
computing their components using definition (\ref{comp}), which gives
\beq \label{c-comp}
\begin{array}{ll}
(C_- \wedge l)_v = (A_-)_n \, ,  &(C_-\wedge n)_v = - (A_-)_l + \nm (A_0)_\mb \, , \\
(C_- \wedge m)_v = - (A_-)_\mb \, , &(C_-\wedge \mb)_v = (A_-)_m - \nm (A_0)_n \, , \medskip \\
(C_+ \wedge l)_v = \nm (A_0)_m - (A_+)_n \, , &(C_+\wedge n)_v = (A_+)_l \, , \\
(C_+ \wedge m)_v = - \nm (A_0)_l + (A_+)_\mb \, , \qquad &(C_+\wedge \mb)_v = - (A_+)_m, \medskip \\
(C_0 \wedge l)_v = \nm (A_-)_m \, ,  &(C_0 \wedge n)_v = \nm (A_+)_\mb \, , \\
(C_0 \wedge m)_v = - \nm (A_-)_l \, , &(C_0 \wedge \mb)_v = - \nm (A_+)_n \, ,
\end{array}
\eeq
and then solving this linear system of equations, with the result
\beq \label{a-comp}
\begin{array}{ll}
(A_-)_l = - \nm^{-1} (C_0 \wedge m)_v \, , &(A_-)_n= (C_- \wedge l)_v \, , \\
(A_-)_m = \nm^{-1} (C_0 \wedge l)_v \, , &(A_-)_\mb = - (C_- \wedge m)_v \, , \medskip \\
(A_+)_l = (C_+ \wedge n)_v \, , &(A_+)_n = - \nm^{-1} (C_0 \wedge \mb)_v \, , \\
(A_+)_m = - (C_+ \wedge\mb)_v\, , &(A_+)_\mb = \nm^{-1} (C_0 \wedge n)_v \, ,
\medskip \\
(A_0)_l = \nm^{-2} (C_0 \wedge n)_v - \nm^{-1} (C_+ \wedge m)_v \, , &(A_0)_n =
\nm^{-2} (C_0 \wedge l)_v - \nm^{-1} (C_- \wedge \mb)_v \, , \\
(A_0)_m = - \nm^{-2} (C_0 \wedge \mb)_v + \nm^{-1} (C_+ \wedge l)_v \, , \ \
&(A_0)_\mb = - \nm^{-2} (C_0 \wedge m)_v + \nm^{-1} (C_- \wedge n)_v \, .
\end{array}
\eeq
Here, the subscripts $l$, $n$, $m$, and $\mb$ indicate the corresponding
components in the development over the basis one-forms (\ref{frame}).

Using (\ref{db-comp}) and (\ref{cs}), we finally have
\beq\label{a}
\begin{array}{c}
\displaystyle A_- = - \frac{1}{\sqrt2\, r \nm g}\,  m \, , \qquad
A_+ = - \frac{1}{\sqrt2\, r \nm g}\, \mb \, , \medskip \\
\displaystyle A_0 = \frac{1}{\sqrt2\, g} \left[ {(r \nm f)' \over r \nm f} - {1
\over r \nm^2} \right] (l + n) - {\cot\theta \over \sqrt2\, r}\, (m - \mb) \, .
\end{array}
\eeq
This is our final expression for the spin coefficients. The usual metric case
is obtained by setting $\nm = 1$.

\subsection{Field equations}

Now we are in a position to derive and solve the field equations of our theory
which are analogs of (vacuum) Einstein equations in GR.  The vacuum field
equations are obtained by varying action (\ref{action}) with respect to $B$:
\beq \label{field}
F + (\Psi + \phi\, {\rm Id}) B = 0 \, .
\eeq
The components of the curvature of connection (\ref{a}) are given by
\beq
F_- = dA_- + A_- \wedge A_0 \, , \quad F_+ = dA_+ + A_0 \wedge A_+ \, , \quad
F_0 = dA_0 + 2A_- \wedge A_+ \, .
\eeq
Using the explicit spinor form (\ref{psi}) for $\Psi$, we get
\beq
(\Psi + \phi\, {\rm Id} )B = \tlX_- (\beta + \phi) B_- + \tlX_+ (\beta + \phi)
B_+ + \tlX (\phi - 2\beta) B_0 \, .
\eeq
Thus, we can write (\ref{field}) in components:
\beq\label{eqs-1}
\begin{array}{l}
dA_- + A_- \wedge A_0 + (\beta+\phi) B_- = 0 \, , \smallskip \\
dA_+ + A_0 \wedge A_+ + (\beta + \phi) B_+ = 0 \, , \smallskip \\
dA_0 + 2A_- \wedge A_+ + (\phi - 2\beta) B_0 = 0 \, .
\end{array}
\eeq
All computations are straightforward in view of (\ref{bs}) and (\ref{a}). Each
of the first two equations in (\ref{eqs-1}) gives rise to the two differential
equations
\beq\label{first}
{f_*' \over r g_*^2 f_*} + \left(1 - {1 \over \nm^2} \right) {1 \over r^2
g_*^2} = - {g_*' \over r g_*^3} = \beta + \phi \, ,
\eeq
while the last equation in (\ref{eqs-1}) gives two additional equations
\beq \label{second}
{\nm^2 \over f_* g_*} \left[ {f_* \over r g_*} \left( 1 - {1 \over \nm^2} + {r
f_*' \over f_*} \right) \right]' = {1 \over r^2} \left( {1 \over g_*^2 - 1 }
\right) = \phi - 2 \beta \, ,
\eeq
where
\beq
f_* := \nm f \, , \qquad g_* := \nm g \, .
\eeq

In view of (\ref{c}) and by virtue of the Bianchi identity $\cD F = 0$, where
$\cD$ is the $A$-covariant derivative, only three of the four equations
(\ref{first}) and (\ref{second}) are independent (this can be verified
directly), and the system of three independent equations can be presented in
the following convenient form:
\beq \label{system}
\left( 2 - \phi_\beta \right) \beta' = - {6 \beta \over r} \, , \qquad g_*^{-2}
= 1 - (2 \beta - \phi) r^2 \, , \qquad { (f_* g_*)' \over f_* g_*} = {3
\phi_\beta \over r (2 - \phi_\beta) } \, .
\eeq

\section{Solution and analysis}
\label{sec:solution}

In this section, we proceed to solving the main system of equations (\ref{c}),
(\ref{system}), which completely determine the two-form $B$ given by (\ref{b})
and the one-form $A$ given by (\ref{a}).

\subsection{Corrections to the metric case}

If our solution happens to be in the regime where $|\phi_\beta| \ll 1$, we have
$\nm^2 \approx 1$, which is thus an approximate ``metricity'' regime.
Equations (\ref{system}) in this approximation lead to the
Schwarzschild--(anti)-de~Sitter form for the functions $f$ and $g$ in metric
(\ref{metric}):
\beq \label{sds}
\beta = {r_s \over 2 r^3} \, , \qquad f^2 = g^{-2} = 1 - {r_s \over r} + \phi_0
r^2 \, ,
\eeq
where $\phi_0$ is the value of the almost constant function $\phi(\beta)$ in
this domain, and $r_s$ is the Schwarzschild radius, which appears as the
integration constant of the solution of the first equation in (\ref{system}).
The constant $\phi_0$ corresponds to the effective cosmological constant:
$\phi_0 = - \Lambda / 3$.

The approximate solution (\ref{sds}) will, in particular, hold in the
asymptotic region of large $r$ if $\phi$ is an analytic function of its
arguments $\tr \left( \Psi^2 \right)$ and $\tr \left( \Psi^3 \right)$ in the
neighborhood of zero. In this case, as can be seen from (\ref{ab}),
$\phi(\beta)$ admits an expansion in powers of $\beta^2$ and $\beta^3$, and the
leading term in this expansion for small $\beta$ is
\beq\label{phi-leading}
\phi(\beta) = \phi_0 \pm \ell^2 \beta^2 + {\cal O} \left(\beta^3 \right)\, ,
\eeq
where $\ell$ is a constant of dimension length.  Thus, we have
\beq
\phi_\beta \approx \pm 2 \ell^2 \beta \approx \pm {\ell^2 r_s \over r^3}
\eeq
so that
\beq
|\phi_\beta| \ll 1 \quad \mbox{for} \quad r \gg \left( \ell^2 r_s \right)^{1/3}
\, .
\eeq

By solving (\ref{system}) perturbatively in the small parameter $\ell^2 r_s /
r^3$, one can obtain the next correction to solution (\ref{sds}) in the domain
of large $r\,$:
\beq \label{ads-cor}
\beta = {r_s \over 2 r^3} \left( 1 \pm {\ell^2 r_s \over 2 r^3} \right) \, ,
\quad g_*^{-2} = 1 - {r_s \over r} \left( 1 \pm {\ell^2 r_s \over 4 r^3}
\right) + \phi_0 r^2 \, , \quad f_*^2 = g_*^{-2} \left( 1 \mp {\ell^2 r_s \over
r^3} \right) \, .
\eeq
According to (\ref{c}), the value of the ``nonmetricity'' parameter $\nm^2$ in
this approximation is given by
\beq \label{c-cor}
\nm^2 = 1 \mp {3 \ell^2 r_s \over 2 r^3} \, .
\eeq

Thus, we can see that, under the assumption of regularity of the function $\phi
\left[\tr \left( \Psi^2 \right), \tr \left( \Psi^3 \right)\right]$ in the
neighborhood of $\Psi = 0$, the solution in the asymptotic region $r \to
\infty$ tends to the metric form (the ``nonmetricity'' parameter $\nm$ rapidly
tends to zero) asymptotically describing a space of constant curvature. It is
in this sense that an analog of the Birkhoff theorem holds in the theory under
consideration.

If $\ell \ll r_s$, then the approximate regime (\ref{ads-cor}) and
(\ref{c-cor}) is valid up to the ``horizon'' $r = r_h$, determined by the
condition $f_*^2 = g_*^{-2} = 0$. At the ``horizon,'' the function $c$ remains
finite, and the function $g_*$, hence, also $g$, diverges.  As in the metric
case, this can be regarded as a coordinate singularity which can be removed by
choosing new time and radial coordinates. In this way, one can pass to the
``black hole'' region $r < r_h$, in which the functions $f_*^2$ and $g_*^{-2}$
change sign and become negative.  Somewhere in this region, the condition
$|\phi_\beta| \ll 1$ (or, equivalently, $\ell^2 r_s / r^3 \ll 1$) may cease to
be valid, and the solution becomes strongly non-metric.  To see what can happen
in this region, we consider general solution.

\subsection{General solution}

First of all, one can note that the first and third equations in (\ref{system})
are singular at the point where $\phi_\beta = 2$.  One can remove this
singularity by passing to a new radial coordinate. The value of $\beta$ itself
can be chosen as such a coordinate. Doing this, one can rewrite the system of
equations (\ref{system}) in terms of this new coordinate:
\beq \label{new}
{d \log r \over d \beta} = {\phi_\beta - 2 \over 6 \beta} \, , \qquad g_*^{-2}
= 1 - (2 \beta - \phi) r^2 \, , \qquad { d \log (f_* g_*) \over d \beta } = -
{\phi_\beta \over 2 \beta } \, .
\eeq
This system is nonsingular and can easily be integrated.  The function $\nm^2$,
as usually, is given by (\ref{c}).  Note that $\beta$ has a physical meaning
being a scalar characterizing the field $\Psi$ according to (\ref{ab}) and
(\ref{psi}).

The metric (\ref{metric}) in the new coordinates $(t, \beta)$ is written as
\ber \label{metric-new}
ds^2 = \left( 1 + \phi_\beta \right) \left[ \left( 1 - {\phi_\beta \over 2}
\right)^{-1} f_*^2 dt^2 - \left( 1 - {\phi_\beta \over 2} \right) \left[ {
r(\beta) \over 3 \beta } \right]^2 g_*^2 d\beta^2 \right]
\nonumber \\
{} - r^2 (\beta) \left( d \theta^2 + \sin^2 \theta\, d\phi^2 \right) \, ,
\eer
where we have used (\ref{c}) and the first equation of (\ref{new}).

Metric in this form indicates that the points where $\phi_\beta = 2$,
corresponding to $\nm^2 = 0$, are hypersurfaces across which the coordinates
$t$ and $\beta$ change their space-time roles.  At these hypersurfaces, metric
(\ref{metric-new}) is degenerate. One can see that our basic fields remain
finite at such points. Indeed, in the new coordinates $(t, \beta)$, the forms
$B_\pm$ are proportional to the functions $\nm f = f_*$ and $\nm g = g_*$,
which remain finite and nonzero at these points, and the potentially dangerous
$l \wedge n$ component in $B_0$ is also finite:
\beq \label{ln}
l \wedge n = f g dt \wedge dr = {dr \over d\beta} \nm^{-2} f_* g_* dt \wedge d
\beta = - { r \over 3 \beta} \left( 1 + \phi_\beta \right) f_* g_* dt \wedge d
\beta \, ,
\eeq
where again we have used (\ref{c}) and the first equation of (\ref{new}).  The
same is true for the spin coefficients $A_\pm$ given by (\ref{a}), which are
proportional to $g_*^{-1}$. The potentially dangerous part of the spin
coefficient $A_0$ is finite as well:
\beq
\frac{1}{\sqrt2\, g} \left[ {(r \nm f)' \over r \nm f} - {1 \over r \nm^2}
\right] (l + n) = {f_* \over r g_*} \left({r f'_* \over f_*} + 1 - {1 \over
\nm^2} \right) dt = r f_* g_* (\beta + \phi) dt \, ,
\eeq
where we have used equation (\ref{first}).  Due to Eq.~(\ref{field}), the
components of the curvature $F$ are also finite.

At a point where $\phi_\beta = -1$, the quantity $\nm^{-2}$ turns to zero.
However, by similar reasoning, one can see that all components of $A$ and $B$
remain finite in the coordinates $(t, \beta)$. Therefore, the general solution
is nonsingular at this point as well.  The metric in the form
(\ref{metric-new}) is degenerate, but this is not surprising as we are dealing
with an intrinsically nonmetric theory, and the ``nonmetricity'' parameter
$\nm$ at this point is infinite.

As we already noted in the previous subsection, the condition $g_*^{-2} = 0$,
hence $f_*^2 =0$, corresponds to a horizon in metric (\ref{metric-new}).  This
is a coordinate singularity which can be removed by passing to a Kruskal-like
coordinate system.

For an illustration, let us consider the function $\phi(\beta)$ exactly in the
form
\beq \label{example}
\phi(\beta) = \phi_0 \pm \ell^2 \beta^2 \, ,
\eeq
with two possible signs.  In this case, the solution is
\beq \label{inter}
r^3 (\beta) = { r_s \over 2\beta}\, e^{\pm \ell^2 \beta } \, , \qquad g_*^{-2}
= 1 - \left( 2 \beta \mp \ell^2 \beta^2 - \phi_0 \right) r^2 (\beta) \, ,
\qquad f_* g_* = e^{\mp \ell^2 \beta } \, ,
\eeq
and the function $\nm^2$ is given by
\beq \label{c-inter}
\nm^2 = {1 \mp \ell^2 \beta \over 1 \pm 2 \ell^2 \beta } \, .
\eeq
For $\ell^2 \beta \ll 1$, this corresponds to the approximate solution
(\ref{ads-cor}), (\ref{c-cor}).

With the upper sign in (\ref{example}), from the first equation of
(\ref{inter}), one can see that there is a minimum value of the radial
coordinate
\beq \label{rmin}
r_m^3 = \frac{e}{2}\, \ell^2 r_s \, ,
\eeq
which is precisely the point where $\phi_\beta = 2$.

In the case of the lower sign in (\ref{example}), the function $\beta (r)$ is
monotonic, and $\beta \to \infty$ as $r \to 0$, which resembles the behavior
inside a classical black hole in GR. The point $\phi_\beta = 2$ is absent in
this case, but we have a singularity in metric (\ref{metric-new}) at the point
$\ell^2 \beta = 1/2$ corresponding to the condition $\phi_\beta = -1$.  As we
noted above, the fundamental fields $A$ and $B$ are regular at this point.

\subsection{Minimal black hole}

For both signs in (\ref{example}), there is a minimum value of the
``Schwarzschild radius'' $r_s = r_*$ for which the black-hole horizon exists.
The mechanism, however, is different for the two signs. For the upper sign
(corresponding to positive $\phi$), there is a minimum value (\ref{rmin}) that
the coordinate $r$ can take. This value of $r$ is also the minimum of the
function $g_*^{-2}$. For the horizon to exist, this minimum value should be
negative. Neglecting, for simplicity, the value of $\phi_0$, we can translate
it into the condition
\beq \label{minplus}
r_s > {2 \over e}\, \ell
\eeq
for the Schwarzschild horizon to exist.  If this condition is violated, then,
instead of the Schwarzschild-like horizon, one has a naked surface of
``non-metricity,'' which we describe in the following subsection.

For the lower, negative sign in (\ref{example}), the function $g_*^{-2}$
similarly has a minimum at $\beta=1/\ell^2$. The condition that this minimum is
smaller than zero translates into the condition
\beq \label{minminus}
r_s > {2 e \over 3^{3/2}}\, \ell
\eeq
for the existence of horizon.

\subsection{Conformal structure}

In the modified theory of pure gravity described by action (\ref{action}), the
notion of a unique distinguished metric is replaced by the conformal class of
metrics with respect to which the two-form $B$ is self-dual, with metric
(\ref{metric}) being a representative of this class. A distinguished physical
metric from this class may arise only after one considers gravitational
interaction of matter and radiation.  Several potential candidates for such a
metric can be envisaged at this level.  Given the spinor two-form $B$, one can
consider the Urbantke metric $g^{\rm U}_{\mu\nu}$, defined by the relation
\cite{Urbantke}
\beq \label{urbantke}
\sqrt{|g^{\rm U}|}\, g^{\rm U}_{\mu\nu} = \frac13
\epsilon^{\alpha\beta\gamma\delta} \tr\left( B_{\mu\alpha} B_{\beta\gamma}
B_{\delta\nu} \right) \, ,
\eeq
where the trace is taken with respect to the spinor indices.  It is easy to see
that the Urbantke metric is related to the metric defined by (\ref{metric}) by
the conformal factor $\nm^{2/3}$.

Another distinguished conformal factor for the metric is obtained by the
requirement that the quantity
\beq \label{volume}
\frac{1}{3} \tr \left(B \wedge B \right)
\eeq
coincide with the volume element defined by the new metric.  The metric line element
that arises this way is given by
\beq\label{metric-V}
ds^2_V = \left(\frac{2 \nm^2 + 1}{3}\right)^{1/2} \, ds^2 = \left( 1+\phi_\beta
\right)^{- 1/2} \, ds^2 \, ,
\eeq
where $ds^2$ is the line element given by (\ref{metric}).

In the metric case, where $\nm = 1$, all such definitions coincide with our
``canonical'' metric (\ref{metric}).

As we noted, before one considers coupling of our theory of gravity to matter,
it is impossible to distinguish any of the listed possibilities for the metric
as being the physical one.  However, a rather natural requirement that the
coupling of matter degrees of freedom to $B$ is at most quadratic in $B$ (the
coupling of YM fields satisfies this requirement) favors the metric defined by
(\ref{metric-V}). Indeed, the usual matter coupling to the Urbantke metric
(\ref{urbantke}) would be non-polynomial in the $B$ field, which is undesirable
for many reasons.

In spite of the mentioned ambiguities, the presence of a distinguished
conformal class of metrics, with respect to which the two-form field $B$ is
self-dual, allows us to speak about the conformal structure of the obtained
solution.  This conformal structure can be shown to have physical meaning
reflecting the geometry of propagation of light.

The conformal structure of our black-hole solution in the $(t, \beta)$
coordinate plane will strongly depend on the shape of our basic function $\phi
(\beta)$. However, its key details are easy to understand by looking at the
form of metric (\ref{metric-new}) in the $(t, \beta)$ coordinates.  The metric
described by (\ref{metric-new}) has the following types of critical surfaces
defined by special positions in the $\beta$ coordinate:
\begin{enumerate}
\item \label{sin1}
The point where $g_*^{-2} = 0$, hence, also $f_*^2 = 0$ [the product $f_* g_*$
is positive and finite for finite $\beta$ in view of the last equation in
(\ref{new})]. It defines a null horizon, analogous to the Schwarzschild
horizon, separating different space-time regions in the black-hole solution.
This is a coordinate singularity in $(t, r)$ or $(t, \beta)$ coordinates, which
can be removed by proceeding to Kruskal-like coordinates.
\item \label{sin2}
The point where $\phi_\beta = 2$, or $\nm^2 = 0$.  By differentiating the
function $g_*^{-2}$, one can verify that this is a critical point (maximum or
minimum) of this function, as well as of the radius $r(\beta)$ as a function of
$\beta$.  Hence, typically, this critical point will not coincide with the
previous one, where $g_*^{-2}$ vanishes.  This is a position of true
singularity in the class of metrics (\ref{metric-new}). However, as we said
before, the solution in our basic fields $A$ and $B$ can well be extended
beyond this surface. What happens at this point in terms of our basic two-form
field $B$ is that the components $B_\pm$ and $B_0$ no longer span a subspace in
the space of two-forms which is self-dual in any metric. One can notice, for
example, that $\nm g dr = g_* \left( dr / d \beta \right) d \beta = 0$ at this
point, so the one-forms
\beq
\nm n = \nm l = {1 \over \sqrt{2}} f_* dt \, ,
\eeq
and the anti-self-dual two-forms
\beq
\tilde B_- = \nm\, m \wedge n \, , \qquad \tilde B_+ = \nm\, l \wedge \mb \, ,
\eeq
coincide with their self-dual counterparts $B_-$ and $B_+$, respectively. This means
precisely that there is no metric with respect to which the tripple $B_\pm, B$ is self-dual.
\item \label{sin3}
The point where $\phi_\beta = - 1$, or $\nm^{-2} = 0$.  This is another
singularity in metric (\ref{metric-new}).  At this point, we have $l \wedge n =
0$ in view of (\ref{ln}), and the two-form $B_0$ coincides with its
anti-self-dual counterpart $\tilde B_0\,$:
\beq
B_0 = l \wedge n - m \wedge \mb = - \left(l \wedge n + m \wedge \mb \right) =
\tilde B_0 \, .
\eeq
Again, the fields $A$ and $B$ are well-behaved at this point, so we can cross
it and proceed to a neighboring space-time region.
\end{enumerate}
\begin{figure}
\centerline{\psfig{figure=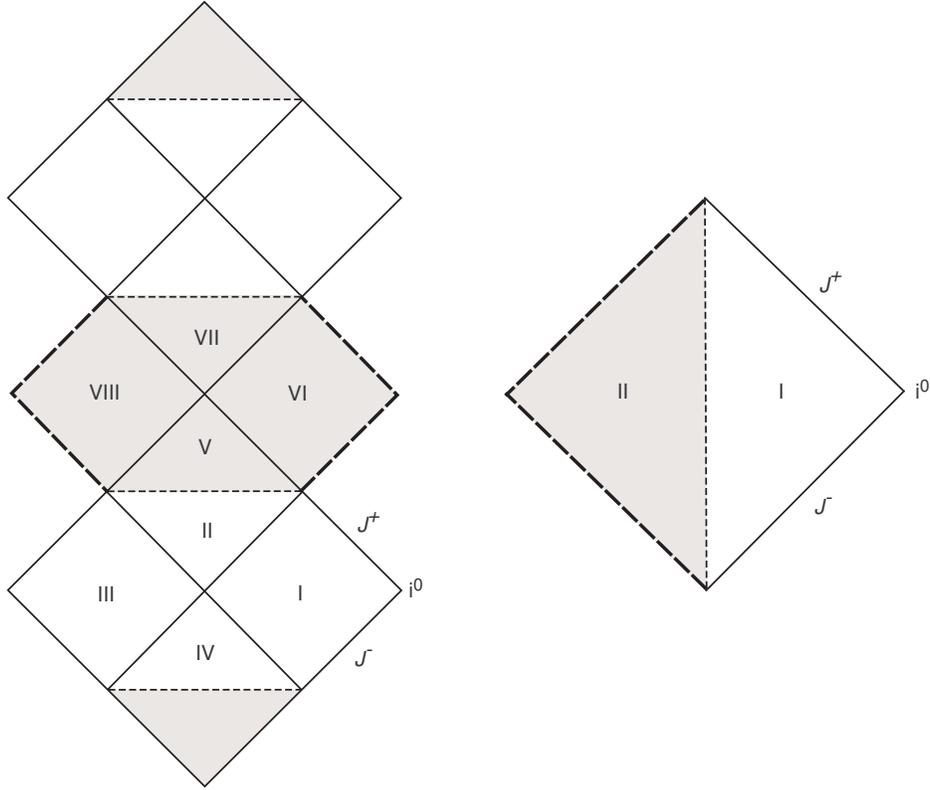,width=0.77\textwidth,angle=0} }
\caption{Conformal diagram of the spherically symmetric solution with the
function $\phi(\beta) = \ell^2 \beta^2$ and with the Schwarzschild radius
satisfying (left image) and violating (right image) condition (\ref{minplus}).
Different regions are numbered in such a way that the coordinate $t$ is
timelike in the odd regions, and spacelike in the even ones. Solid lines
between regions indicate Schwarzschild-like horizons of type \ref{sin1}, at
which $g_*^{-2} = 0$. Thin dashed lines indicate singular surfaces of type
\ref{sin2}, where metric ceases to exist.  The flow of time is vertical in the
white regions and changes to horizontal in the grey regions. Thick dashed lines
indicate the true singularity, where $r = 0$ and $\beta = \infty$. The
configuration on the left image extends periodically and indefinitely upward
and downward. We are living in one of the regions of type I; the asymptotic
spatial infinity in this region is denoted by $i^0$, and the future and past
null infinities are denoted by ${\cal J}^+$ and ${\cal J}^-$, respectively.
\label{fig:conf}}
\end{figure}

As an example, in Fig.~\ref{fig:conf} we have pictured the conformal diagram
obtained for the function $\phi (\beta) = \ell^2 \beta^2$, which has only
critical surfaces of type \ref{sin1} and \ref{sin2}.  The Schwarzschild-like
horizons are pictured by solid lines, while the dashed lines correspond to
horizons of type \ref{sin2}. The radius $r$ has an absolute minimum value, in
this case given by (\ref{rmin}), which is reached precisely at horizons of type
\ref{sin2}. As $\beta \to \infty$, which corresponds to $r \to \infty$, one
approaches a ``singularity'' indicated by thick dashed lines.

Since, in the absence of matter couplings, the physical metric is not
specified, the issue of ``geodesic completeness'' of solution is not well
defined in the purely gravitational theory under consideration. Nevertheless,
it is interesting and instructive to see whether the metric defined by
(\ref{metric}) is geodesically complete as one approaches the singularity
$\beta \to \infty$ in the regions of type VI and VIII. It is easy to see that
null geodesic completeness in such regions is equivalent to the divergence of
the integral
\beq \label{integral}
\int\limits^\infty {1 + \phi_\beta \over \beta} f_* g_* r (\beta) d \beta \, .
\eeq
Thus, solution (\ref{inter}), (\ref{c-inter}) with the upper sign has
convergent integral (\ref{integral}), hence, the metric (\ref{metric}) defined
by this solution is not null geodesically complete.  However, it is timelike
geodesically complete. Indeed, in the regions in question, $\beta$ plays the
role of the time coordinate since $\nm^2\to -1/2$ as $\beta\to\infty$. To the
leading order $g_*^{-2}\sim \ell^2 \beta^2 r^2(\beta)$ and $dr \sim (\ell^2 /
3) r(\beta) d\beta$. This gives $g^2 dr^2 \sim - 2 \ell^2 d\beta^2/ 9 \beta^2$,
and the proper time to infinity $\beta\to\infty$ is logarithmically divergent.

For the lower sign in
(\ref{inter}), (\ref{c-inter}), integral (\ref{integral}) is divergent;
therefore, the corresponding metric (\ref{metric}) is null geodesically
complete in the region $\beta \to \infty$, or $r \to 0$.  However, it is not
timelike geodesically complete since the timelike distance to the singularity
$\int g dr \propto \int dr$ is finite (the coordinate $r$ is timelike and $r \to
0$ in this case).

The described behavior makes the ``singularity'' at $\beta\to\infty$ rather an
interesting place. We have depicted it as a null surface because, in some
respects, it is reminiscent of the usual null infinity. The reader should,
however, keep in mind that the detailed structure of this ``singularity'' is
quite unusual, and strongly depends on the details of the behavior of the
function $\phi$ as $\beta\to\infty$. In contrast, the structure of the
``non-metric horizons'' at $\phi_\beta=2,-1$ is universal. It is also worth
emphasizing that the singularity at $\beta\to\infty$ is located ``outside'' of
the black hole on the left image in Fig.~\ref{fig:conf} in the sense that
another Schwarzschild-like horizon has to be crossed to reach it. It is in this
sense that the theory under consideration ``resolves'' the singularity inside
the black hole.

In the examples considered above, we have assumed that the function
$\phi(\beta)$ behaves as $\pm \ell^2 \beta^2$ as $\beta\to\infty$. Some other
choices are possible and may be interesting to consider. Thus, it is easy to
devise $\phi(\beta)$ so that there is a maximal possible ``curvature'' in the
theory. To achieve this, one can choose $\phi(\beta)$ diverging for some finite
value of $\beta$. A good example of such a function is given by $\phi(\beta) =
x^2/ \ell^2 (1 - x^2)$, where, as before, $x=\ell^2\beta$.  It diverges at the
value $\beta_{max}=1/\ell^2$, which plays the role of maximal curvature in the
theory. The qualitative behavior of the conformal diagram for the spherically
symmetric solution will be the same in this case. The details of the behavior
at the singularity at $\beta_{max}$ will, however, be different. Another
interesting possibility not considered in this paper is that, for Planckian
curvatures, the function $\phi(\beta)$ changes rapidly --- so that, in
particular, one of the non-metric horizons at $\phi_\beta=2,-1$ is reached ---
and approaches a large constant value for $\beta$ much larger than Planckian.
In this scenario, we would get a conformal diagram consisting of elements
similar to the ones described above, with a metric universe with large
cosmological constant on the other side of the black hole. It could be
interesting to study all such possibilities in more detail.

\section{Modification of gravity at different curvatures}
\label{sec:mod}

The theory under consideration leads to a possibility of scale-dependent
modification of gravity of very interesting nature.

Defining the {\em effective\/} Schwarzschild radius $r_s (\beta)$ by the
relation
\beq
r_s (\beta) = 2 \beta r^3 (\beta) \, ,
\eeq
we can present our main system of equations (\ref{new}) in the form
\beq \label{mod}
{d \log r_s \over d \beta} = {\phi_\beta \over 2 \beta} \, , \qquad g_*^{-2} =
1 - {r_s \over r} + \phi r^2 \, , \qquad { d \log (f_* g_*) \over d \beta } = -
{\phi_\beta \over 2 \beta } \, .
\eeq
Integrating the first and the last equations in (\ref{mod}), we get the
following relations, valid for arbitrary $\beta_1$ and $\beta_2\,$:
\beq \label{mass}
{r_s (\beta_2) \over r_s (\beta_1)} = Z (\beta_1, \beta_2) \, , \qquad { f_*
(\beta_2) \over f_* (\beta_1) } = { g_* (\beta_1) \over g_* (\beta_2) }\,
Z^{-1} (\beta_1, \beta_2) \, ,
\eeq
where
\beq \label{z}
Z(\beta_1, \beta_2) = \exp \left( \int_{\beta_1}^{\beta_2} {\phi_\beta \over 2
\beta} d \beta \right) \, .
\eeq
The first equation in (\ref{mass}) shows that we are dealing with a theory in
which the effective Schwarzschild radius becomes distance-dependent, and the
second equation indicates the presence of an additional redshift/blueshift
factor in the metric.

Assume that there exist two domains in the space of values of $\beta$, in both
of which one has $|\phi_\beta| \ll 1$, hence, both of which are characterized
by the condition of metricity.  Then the Schwarzschild radii will be constant
in the corresponding regions of the radial coordinate and will be related by
(\ref{mass}) in which $\beta_1$ and $\beta_2$ are the representative values of
$\beta$ in the corresponding domains.  The time-time component of the metric,
according to the second relation of (\ref{mass}), will exhibit the relative
redshift factor between these two regions
\beq \label{redshift}
{ g_{00}^{(2)} \over g_{00}^{(1)} } = { g_{rr}^{(1)} \over g_{rr}^{(2)} }\,
Z^{-2} \, .
\eeq
The cosmological constant, in general, will also be renormalized.

As an illustration for these effects, we consider an explicit example, which
is defined by the function
\beq \label{phi-log}
\phi(\beta) = \phi_0 - \phi_1 \log \left( 1 + x^2 \right) \, , \qquad x :=
\ell^2 \beta \, .
\eeq
The derivative of $\phi$ is given by the expression
\beq \label{deriv-log}
\phi_\beta = - {2 \alpha x \over 1 + x^2 } \, ,
\eeq
where
\beq
\alpha = \ell^2 \phi_1
\eeq
is a dimensionless parameter.  Function (\ref{phi-log}) has the property that
$|\phi_\beta| \ll 1$ in the two asymptotic regions $x \ll \left(1 + |\alpha|
\right)^{-1}$ and $x \gg 1 + |\alpha|$, so that the theory is approximately
``metric'' in both these regions.

Equations (\ref{new}) can be integrated in this case, with the solution
\beq \label{sol-log}
{\ell^2 r_s \over 2 r^3} = x e^{\alpha \arctan x - {\pi \alpha \over 2}} \, ,
\qquad f_* g_* = e^{\alpha \arctan x - {\pi \alpha \over 2}} \, .
\eeq

We have chosen the integration constants in (\ref{sol-log}) so as to obtain the
standard Schwarzschild solution at small radial distances, $r^3 \ll \left( 1 +
|\alpha| \right)^{-1} \ell^2 r_s$, where $x \gg \left(1 + |\alpha| \right)$. At
these distances, we can expand the right-hand sides of solution (\ref{sol-log})
to obtain
\beq
{\ell^2 r_s \over 2 r^3} = x \left[ 1 - {\alpha \over x} + {\cal O}
\left(\alpha x^{-3} \right) \right] \, , \qquad f_* g_* = 1 - {\alpha \over x }
+ {\cal O} \left( \alpha x^{-3} \right) \, , \quad x \gg \left(1 + |\alpha|
\right) \, .
\eeq
We also have
\beq
\nm^2 = 1 + {3 \alpha \over x} + {\cal O} \left(\alpha x^{-2} \right) \, ,
\quad x \gg \left(1 + |\alpha| \right) \, .
\eeq
Collecting all the terms, we obtain the leading contribution to the $g_{00}$
coefficient in metric (\ref{metric}) and $\tilde g_{00}$ coefficient in the
Urbantke and ``volume'' metric (which coincide in this approximation):
\beq \label{small-log}
\begin{array}{rcl}
g_{00} &=& \displaystyle 1 - {r_s \over r} + \left[ \phi_0 - 2 \phi_1 \left( 1
+ \log { \ell^2 r_s \over 2 r^3 } \right) \right] r^2 - {10 \alpha  r^3 \over
\ell^2 r_s} \, ,  \bigskip \\
\tilde g_{00} &=& \displaystyle 1 - {r_s \over r} + \left[ \phi_0 - 2 \phi_1
\left( 1 + \log { \ell^2 r_s \over 2 r^3 } \right) \right] r^2 - {8 \alpha  r^3
\over \ell^2 r_s} \, , \quad r^3 \ll { \ell^2 r_s \over 1 + |\alpha| } \, .
\end{array}
\eeq

At large radial distances $r^3 \gg \left(1 + |\alpha| \right) \ell^2 r_s$,
where $x \ll \left(1 + |\alpha| \right)^{-1}$, we have
\beq
{\ell^2 r_s \over 2 r^3} = e^{-{\pi \alpha \over 2}} x \left[ 1 + \alpha x +
{\cal O} \left(\alpha x^3 \right) \right] \, , \qquad f_* g_* = e^{-{\pi \alpha
\over 2}} \left[ 1 + \alpha x + {\cal O} \left( \alpha x^3 \right) \right] \, ,
\eeq
\beq
\nm^2 = 1 + 3 \alpha x + {\cal O} \left(\alpha x^2 \right) \, , \quad x \ll { 1
\over 1 + |\alpha| } \, .
\eeq
In this case, we obtain the following leading contribution to the corresponding
metrics, up to terms of order $\ell^2 r_s / r^3$:
\beq \label{large-log}
\begin{array}{rcl}
g_{00} &=& \displaystyle e^{- \pi \alpha} \left[ 1 - e^{\pi \alpha \over 2}
{r_s \over r} + \phi_0 r^2  - e^{\pi \alpha \over 2} {\alpha \ell^2 r_s \over 2
r^3 } \right] \, , \bigskip \\
\tilde g_{00} &=& \displaystyle e^{- \pi \alpha} \left[ 1 - e^{\pi \alpha \over
2} {r_s \over r} + \phi_0 r^2 \right] \, , \quad r^3 \gg \left( 1 + |\alpha|
\right) \ell^2 r_s \, .
\end{array}
\eeq
We note that $\nm^2 \approx 1$ in our approximation in these regions, so the
physical metric is well defined. In fact, the contributions proportional to
$\alpha$ in (\ref{small-log}) and (\ref{large-log}) are small and can be
dropped if $\alpha$ is not very large.

In the intermediate region of radial distances
\beq \label{between}
{1 \over 1 + |\alpha| } \lesssim {r^3 \over \ell^2 r_s } \lesssim 1 + | \alpha
| \, ,
\eeq
the metric coefficient behaves in a complicated way; moreover, the physical
metric is not well defined in this region unless $\alpha \ll 1$.  However, even
without the detailed knowledge of matter couplings to gravity, simply on the
basis of physical continuity, it is obvious that the {\em apparent\/} mass (or
gravitational coupling) will be varying continuously between the two asymptotic
regions. The intermediate region (\ref{between}) is rather extended if
$|\alpha| \gg 1$.

It is interesting to note that the shift in the cosmological constant between the
two asymptotic regions will be absent in a special case where the function $\phi (\beta)$
takes the same value in the corresponding domains of the $\beta$ space.  In spite of
the absence of such a shift, the Schwarzschild radius will still get renormalized. This can be
illustrated by another example:
\beq \label{phi-mod1}
\phi(\beta) = \phi_0 - \phi_1 {x^2 \over 1 + x^3} \, , \qquad x := \ell^2 \beta
\, ,
\eeq
which is a function of both invariants $\tr \left(\Psi^2\right) = 6 \beta^2$
and $\tr \left(\Psi^3\right) = - 6 \beta^3$ in the case $\alpha = 0$ [see
Eq.~(\ref{ab})].  This function is characterized by the property that $\phi
\approx \phi_0$ in both asymptotic domains $x \ll 1$ and $x \gg 1$.  The mass
renormalization exponent in (\ref{z-gen}) is given in this case by the integral
\beq
- \int_0^\infty {\phi_\beta \over 2 \beta}\, d \beta = \alpha
\int\limits_{0}^{\infty} {1 - \frac12 x^3 \over \left( 1 + x^3 \right)^2 } \, d
x = {\pi \alpha \over 3 \sqrt{3}}  \, ,
\eeq
where $\alpha = \ell^2 \phi_1$.

In the case under consideration, we obtain the following approximate
expressions for the metric, respectively, at small and large radial distances:
\ber
g_{00} = 1 - {r_s \over r} + \phi_0 r^2  + 2 \alpha \left( { 2 r^3 \over
\ell^2 r_s} \right)^2 \, , &{}& \nonumber \\
\tilde g_{00} = f^2 = 1 - {r_s \over r} + \phi_0 r^2  + \frac32 \alpha \left( {
2 r^3 \over \ell^2 r_s} \right)^2 \, , &\quad& r^3 \ll \displaystyle { \ell^2
r_s \over 1 + |\alpha| } \, , \label{small-mod1}
\eer
\ber
g_{00} = e^{-{2 \pi \alpha \over 3 \sqrt{3}}} \left[ 1 - e^{\pi \alpha \over 3
\sqrt{3}} {r_s \over r}  + \phi_0 r^2  - e^{\pi \alpha \over 3 \sqrt{3}}
{\alpha \ell^2 r_s \over 2 r^3 } \right] \, , &{}& \nonumber \\
\tilde g_{00} = e^{-{2 \pi \alpha \over 3 \sqrt{3}}} \left[ 1 - e^{\pi \alpha
\over 3 \sqrt{3}} {r_s \over r}  + \phi_0 r^2 \right] \, , &\quad& r^3 \gg
\left( 1 + |\alpha| \right) \ell^2 r_s \, . \label{large-mod1}
\eer
More general set of examples, characterized by arbitrary powers in the region
of large and small curvatures is considered in Appendix~\ref{app:general}. The
above formulas are in correspondence with the generic expressions
(\ref{small-gen}) and (\ref{large-gen}).

\section{Physical effects}
\label{sec:effects}

\subsection{Modified gravity instead of dark matter}

Modifications of gravity at large distances are currently under consideration
as an alternative to the dark-matter phenomenon (see, e.g., \cite{Moffat}).
Our theory can be a viable candidate in this respect.  Indeed, we have seen in
the previous section that the gravitational strength of the central spherically
symmetric body is distance-dependent.  Consider, for definiteness, our example
(\ref{phi-log}). Because the parameter $\alpha$ stands in the exponent of the
renormalized gravitational mass [see Eq.~(\ref{large-log})], one should have
positive $\alpha \sim 1$ in order that the gravitational mass increase with
distance. Then the fundamental length scale $\ell$ should be chosen in such a
way as to ensure that the deviations from the Newtonian behavior begin at a
certain distance from the gravitating body.  A typical representative of such a
situation is a spiral galaxy like our Milky Way, of mass $M_g \sim 10^{11}
M_\odot$, in which deviations from Newton's behavior (flat rotation curves) are
prominent at a distance $r_g \sim 10$~kpc.  This gives us the estimate
\beq \label{ell}
\ell \sim \sqrt{r_g^3 \over r_s} \sim 10~\mbox{Mpc} \, ,
\eeq
where $r_s = 2 GM_g \sim 10^{-2}$\,pc is the Schwarzschild radius associated
with the galaxy mass contained inside the radius $r_g$. Interestingly, this
estimate for $\ell$ roughly corresponds to the scale on which the relative
perturbation $\delta \rho / \rho$ of the density in the universe becomes of
order unity today.

The critical distances $r_c \sim \left(\ell^2 r_s \right)^{1/3}$ for the Sun,
for the Earth, and for an isolated proton are then given by
\beq \label{critical}
r_\odot \sim 2~\mbox{pc} \, , \qquad r_\oplus \sim 2 \times 10^3~\mbox{au} \, ,
\qquad r_p \sim 5~\mbox{mm} \, ,
\eeq
respectively.  The large practical value of the critical radius
(\ref{critical}) even for microscopic bodies such as a proton will result in
the effect that, in most (if not all) practical cases in the given example, the
overall redshift factor (\ref{redshift}) of low-curvature regions will be
unobservable. In other words, the photons (and other particles) will always be
emitted from regions of relatively high curvature (magnitudes of curvature
corresponding to $\ell^2 \beta \gg 1$ in the spherically symmetric case) and
observed also in regions of high curvature, and there is no relative redshift
factor between such regions.

To explain this in more detail, assuming that we have $N$ gravitational sources
separated by distances large compared to their critical scales $r^{(i)}_{c} =
\left(\ell^2 r^{(i)}_{s} \right)^{1/3}$ determined by their small-distance
Schwarzschild radii $r^{(i)}_{s}$, $i = 1,\ldots,N$, we can write the metric in
the region far from all such sources as
\beq \label{far}
g_{00} = 1 - \sum_i {\tilde r^{(i)}_{s} \over r_i}  \, , \qquad r_i \gg
r^{(i)}_c \, , \quad i = 1,\ldots,N \, ,
\eeq
where $r_i$ is the distance from the point of observation to the $i^{\rm th}$
source, and $\tilde r^{(i)}_{s} = Z r^{(i)}_{s}$ is the renormalized
Schwarzschild radius with the universal factor $Z$. We have used the
superposition principle clearly valid in the domain $\ell^2 \beta \ll 1$ and
neglected the contribution from the effective cosmological constant. As we
approach any of the sources within its critical radius, the metric is dominated
by this source and reads
\beq
g^{(i)}_{00} = Z^2  \left( 1 - {r^{(i)}_{s} \over r_i} \right) \, , \quad r_i
\ll r^{(i)}_c \, ,
\eeq
with the same universal redshift factor $Z$.  Because of this universality, a
photon or any other particle traveling from one such source to another one
will experience no additional relative redshift.

Thus, for any sufficiently massive practical observer, only the effect of the
apparent increase of the central mass (or, equivalently, of the gravitational
constant) with distance will be detectable, which asymptotically will result in
the difference between $r_{s}$ and $\tilde r_{s} = Z r_s$. Although the notion
of physical metric may not be well defined at intermediate distances
(\ref{between}), it is clear that the effect of apparent increase of the
gravitational mass of the central body should be continuous with distance;
hence, the value of the effective mass in our example will monotonically
interpolate between its asymptotic magnitudes $r_s$ and $Z r_s$ as one moves
away from the gravitating body. This potentially can be used to explain the
effect of missing gravitating mass in the universe.

\subsection{Pioneer anomaly}

It is interesting to see whether the theory under consideration might be able
to explain the observed anomalous acceleration of the Pioneer spacecraft
$a_{\rm P} \simeq 8 \times 10^{-10}$~m/s$^2$ \cite{Pioneer}. To explain the
alleged missing mass in galaxies one has to take the function $\phi(\beta)$
such that the effective gravitating mass increases with the distance. This
would imply the presence of anomalous acceleration directed towards the Sun,
potentially matching the observed value.

To derive the exact observable result, we need to consider concrete realistic
functions $\phi(\beta)$ as well as the couplings of matter to the gravitational
degrees of freedom of this theory, which will possibly distinguish one of the
effective metrics from the whole conformal class. Some preliminary
order-of-magnitude estimates can already be made irrespective of such details.
The leading terms in non-metric corrections to the acceleration at small
distances are expected to be of the order
\beq \label{anom}
a_0 \simeq {\alpha r^2 \over \ell^2 r_s } \, ,
\eeq
as can be seen from the last terms on the right-hand side of (\ref{small-gen})
or (\ref{small-alt}) of Appendix~\ref{app:general} for the simplest generic
case $p_2 = 1$ [see also the concrete example (\ref{small-log})]. (To avoid
confusion, we remind the reader that the speed of light was set to unity
throughout this paper.) For the value of $\ell/ \sqrt{\alpha} \simeq 20$~Mpc,
the Pioneer acceleration $a_{\rm P} \simeq 8 \times 10^{-10}$~m/s$^2$ will be
reached at a distance from the Sun $r = 20$~au.  This value of $\ell/
\sqrt{\alpha}$ is of the same order as that which we obtained in the
explanation of the missing mass in galaxies [see Eq.~(\ref{ell})].  Thus,
together with accounting for the missing mass in galaxies, the present theory
may also be able to explain this observed anomaly in a uniform way.

Because of the quadratic dependence on the distance in (\ref{anom}), the
anomalous acceleration of the inner planets of the solar system will be well
below the observational limits.  However, the theoretical prediction for the
outer planets based on the simple estimate (\ref{anom}) seems to disagree with
the observational constraints, as can be seen from Table~\ref{table}. The
observational constraints for Uranus and Neptune, if proved to be valid, are
also inconsistent with the general explanation of the Pioneer anomaly as a
modification of gravity \cite{Sanders}.
\begin{table}
{
\begin{tabular}{|l|c|c|c|}
\hline \ \ Object & \ \ Distance (au) \ \ & \ \ $a_{\rm P}^{\rm theor}$ $\left( 10^{-10}\,
\mbox{m/s$^2$} \right)$ \ \ &  \ \ $a_{\rm P}^{\rm obs}$ $\left( 10^{-10}\,
\mbox{m/s$^2$} \right)$ \ \ \\
\hline \hline
\ \ Mercury \ \  & 0.39 & 0.003 & 0.04 \\
\hline
\ \ Icarus & 1.08 & 0.02 & 6.3\\
\hline
\ \ Mars & 1.52 & 0.04 & 0.1\\
\hline
\ \ Jupiter & 5.2 & 0.5 & 0.12\\
\hline
\ \ Uranus & 19.2 & 7 & 0.08$^*$\\
\hline
\ \ Neptune & 30.1 & 17 & 0.13$^*$\\
\hline
\end{tabular}
} \bigskip \caption{The final column is the upper limit on constant
acceleration determined from planetary orbits and taken from \cite{Sanders}.
The constraints imposed by the orbits of Uranus and Neptune are somewhat
uncertain; for this reason, they are marked with asterisk (see \cite{Sanders}).
The third column is the theoretical anomalous acceleration estimated from
(\ref{anom}) with $\ell / \sqrt\alpha = 20$~Mpc. Theoretical estimates for
Uranus and Neptune, if proved to be valid, significantly exceed the
observational constraints, which are also inconsistent with the Pioneer anomaly
as a modification of gravity \cite{Sanders}.} \label{table}
\end{table}

The anomalous acceleration (\ref{anom}) is obtained for functions $\phi(\beta)$
which have logarithmic asymptotic behavior at relatively large values of
$\beta$, as in example (\ref{phi-log}), which results in metric
(\ref{small-log}) at small distances. If the function $\phi(\beta)$ has
power-law behavior at large values of $\beta$, as in example (\ref{phi-mod1}),
then the modified metric at small distances takes the form (\ref{small-mod1}),
leading to the anomalous acceleration of the order
\beq \label{anom1}
a_0 \simeq {\alpha \over r } \left( {r^3 \over \ell^2 r_s } \right)^2 \, ,
\eeq
which differs from (\ref{anom}) by a small factor $r^3 / \ell^2 r_s$, of the
order $10^{-14}$ for Uranus and Neptune.  In this case, the expected anomalous
acceleration is many orders of magnitude smaller than that observed for the
Pioneer, and satisfies well the solar-system constraints derived from planetary
motion.

Thus we can see that the specific prediction for anomalous accelerations within
the solar system is strongly sensitive to the form of the unknown function
$\phi(\beta)$ in the domain of relevant curvatures $\beta$.  It looks possible
to explain the Pioneer anomaly (if it has gravitational origin) in frames of
the modified theory of gravity under consideration, but a concrete realization
of this requires more work.

\subsection{Non-cosmological redshifts of quasars and gamma-ray bursts}

The idea that high redshifts of quasars are not of cosmological nature, but
rather are caused by some poorly understood physical circumstances, is pursued
by a number of astrophysicists \cite{quasars} (see also \cite{Corredoira} for
critical assessment and \cite{Collin:2006} for a historical review). Similar
considerations exist concerning the high redshifts of gamma-ray bursts
\cite{Burbidge}. One of the problems with this controversial hypothesis is to
propose a viable alternative explanation for such high redshifts. Here we would
like to demonstrate that our theory may be capable of providing such an
explanation.

Suppose that the condition $\phi_\beta > 0$ is satisfied in the range $\beta_1
< \beta < \beta_2$ so that the redshift factor $Z (\beta_1, \beta_2)$ in
(\ref{z}) is bigger than unity. Then, assuming for the moment that the end
points of this range are in the metricity regime $\phi_\beta \ll 1$, and taking
into account that the value of $\beta$ in the corresponding regions is the
value of the Weyl curvature of the physical metric, we conclude that light
emitted from regions of higher Weyl curvature will be additionally redshifted
by the factor $Z (\beta_1, \beta_2) > 1$. The observable redshift $z_{\rm obs}$
of a source with such intrinsic redshift factor $Z$ and with local (peculiar or
cosmological) redshift $z$ will then be given by
\beq \label{zobs}
z_{\rm obs} = Z (1 + z) - 1 \, .
\eeq
This effect can be responsible for the observed high redshifts of quasars and
gamma-ray bursts.

The value of $Z (\beta_1, \beta_2)$ is quite {\em arbitrary\/} depending
exponentially on the behavior of the function $\phi (\beta)$.   As a simple
example, consider the function
\beq \label{phi-qso}
\phi (\beta) = \phi_1 \log \left( 1 + x^2 \right) \, , \qquad x = \ell^2 \beta
\, ,
\eeq
which gives the ``metric'' behavior in the two domains $x \ll 1$ and $x \gg 1$.
The derivative $\phi_\beta$ should also satisfy the condition $\phi_\beta < 2$
in order that we do not encounter a singularity outside the horizon. The
validity of this condition for all $\beta$ leads to the constraint
\beq \label{upperb}
\alpha := {\ell^2 \phi_1 \over 2} < 1 \, .
\eeq

The redshift factor is given by the expression
\beq \label{boundz}
Z  = \exp \left(\int\limits_{\,\,\,x_1 \ll 1}^{x_2 \gg 1} {\phi_\beta (x) \over
2 x} d x \right) \approx e^{\pi \alpha} < 23.14 \, ,
\eeq
where the last upper bound comes from the bound in (\ref{upperb}).  The upper
bound (\ref{upperb}), (\ref{boundz}) for the admissible values with function
(\ref{phi-qso}) is more than enough to explain the redshifts of quasars and
gamma-ray bursts up to $z_{\rm obs} \lesssim 6$.  Indeed, we need $\alpha
\approx 0.62$ to get the highest values of $z_{\rm obs} \approx 6$ for $z = 0$.

According to the described scenario, additional redshift factor $Z$ will be
present in {\em all\/} massive compact objects in which radiation originates in
regions with curvature $\beta \gtrsim \ell^{-2}$, i.e., having the following
relation between the size $R$ and Schwarzschild radius $r_s\,$:
\beq \label{qso}
R^3 \lesssim r_s \ell^2 \, .
\eeq
The free parameters $\ell$ and $\alpha$ should be fitted to model redshifts of
compact objects with strong gravity. Thus, if we take an accretion onto a black
hole as a working model for a quasar, then we will have $R \sim r_s$, and
condition (\ref{qso}) will become $r_s \lesssim \ell$, implying that only black
holes of mass {\em smaller\/} than some value produce redshifted emission.  To
fix the value of $\ell$, one needs to build a detailed model of a quasar, which
obviously goes beyond the scope of this paper.  We only stress that now it has
to be built under the assumption that quasars are not superluminous objects
situated at very large distances but rather are compact massive objects of
average absolute luminosity in our proximity.  Also note that the gravitational
redshift in our theory is different by nature to the usual gravitational
redshift from a massive body in GR.  This is clear from the observation that
our effect involves a length scale $\ell$ in (\ref{qso}) and, therefore, can
take place at arbitrary Weyl curvatures, for example, at which the usual
redshift would be negligible.  The redshift itself depends on another free
constant $\alpha$, as can be seen from (\ref{boundz}).  Moreover, our effect in
a wide range of curvature values around the central body can be almost
independent of the value of curvature itself. Hence, the usual constraints
\cite{Greenstein:1964} that rule out the gravitational redshift of quasars in
general relativity have to be revised in the theory under investigation.

There are two possible sources of the observed scatter in the values of
redshifts.  The redshift factor $Z (\beta_1, \beta_2)$ will depend on the
actual curvature $\beta_2$ of the region where radiation is formed. We remember
that the effective physical metric is not yet defined in the region where
$|\phi_\beta| \sim 1$; however, the observable effect of redshift from these
regions is expected to be of the magnitude $Z(\beta_1, \beta_2)$. The spread in
the values of $\beta_2$ will then lead to a spread in the observed redshifts.
Another obvious source of scatter in the redshift values is the overall
cosmological expansion and peculiar motion which contribute to the observable
redshift (\ref{zobs}).  As is clear from (\ref{zobs}), the effect due to
peculiar motion and cosmological expansion is {\em amplified\/} by the factor
$Z$.  For example, if $Z = 2$, then a nearby quasar situated at $z = 0$ will be
observed at $z_{\rm obs} = 1$, while a quasar situated at $z = 0.5$ will be
seen to have $z_{\rm obs} = 2$.

The non-cosmological redshift effect which we consider in this subsection could
also be relevant to neutron stars.  In order that this effect be appreciable,
condition (\ref{qso}) should be satisfied at their surfaces. Considering that
neutron stars have approximately nuclear density, we conclude that the ratio
$R^3 / r_s$ (where $R$ is their radius) is approximately constant in such
objects, and condition (\ref{qso}) then gives a numerical estimate for
$\ell\,$:
\beq \label{ellest}
\ell \gtrsim 30~\mbox{km} \, .
\eeq
With $\ell$ satisfying this condition, radiation from the surfaces of neutron
stars will be additionally effectively redshifted.  This effect, in principle,
may be tested by observations.

Concluding this subsection, we note that the same condition (\ref{ellest}) will
formally imply nontrivial gravitational effects from the region of atomic
nuclei. However, the classical picture of gravity that we considered here may
be not valid in such microscopic regions. This question deserves additional
investigation.

\subsection{Combined scenario}

\begin{figure}
\centerline{\psfig{figure=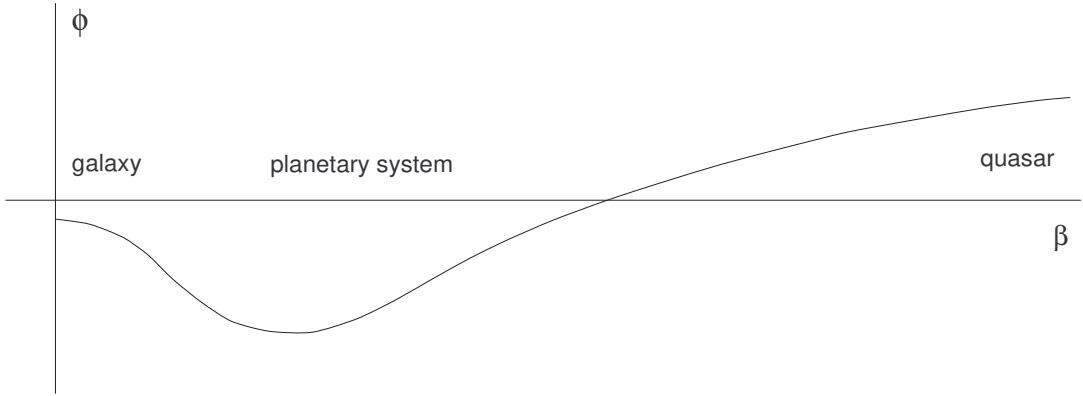,width=0.9\textwidth,angle=0} }
\caption{Typical qualitative shape of the function $\phi(\beta)$ which might
exhibit the three effects discussed in this paper, namely, missing mass in
galaxies, anomalous acceleration in the outskirts of the solar system, and high
redshifts of quasars and gamma-ray bursts.  The titles of astrophysical objects
are placed at the corresponding values of $\beta$, which is roughly the Weyl
curvature of space-time characteristic of these objects. \label{fig:phi}}
\end{figure}

In order to implement simultaneously the described physical effects, the
function $\phi (\beta)$ much be chosen appropriately.  It is clear that we
require the existence of three different domains in the $\beta$ space, with
solar system in the region characterized by the ``metricity'' property
$\phi_\beta \ll 1$. Let us denote representative values of $\beta$ in the order
of increasing by $\beta_1$, $\beta_2$, and $\beta_3$.  The solar-system values
of curvature will correspond to $\beta_2$. We should have $Z (\beta_1, \beta_2)
< 1$ to explain galactic rotation curves, and $Z (\beta_2, \beta_3) > 1$ to
explain high redshifts of quasars and gamma-ray bursts.  The function that may
do the job will look something like
\beq
\phi (\beta) = \phi_0 - \phi_1 \log \left( 1 + x_1^2 \right) + \phi_2 \log
\left( 1 + x_2^2 \right) \, , \qquad x_1 = \ell_1^2 \beta \, , \quad x_2 =
\ell_2^2 \beta \, ,
\eeq
with $\ell_1 \gg \ell_2$ and $\phi_2 > \phi_1$.  The value of $\phi_0$ should
be negative to account for the positive large-scale cosmological constant
$\Lambda = - 3 \phi_0$. Qualitative behavior of the function $\phi(\beta)$ is
shown in Fig.~\ref{fig:phi}.
However, one can propose other functions with the desired properties, and it is
by combined theoretical investigations in quantum gravity and by comparison
with observations that a reasonable function $\phi(\beta)$ may be established.

\section{Summary}
\label{sec:sum}

In this paper, we considered the vacuum spherically symmetric problem in the
theory of gravity described by action (\ref{action}), which modifies the
Pleba{\'n}ski self-dual formulation of general relativity by introducing an
additional function $\phi \left[\tr \left( \Psi^2 \right), \left( \Psi^3
\right) \right]$. In the spherically symmetric case, the two arguments of the
function $\phi$ are not independent, and it becomes a function of only one
effective field $\beta$, defined in (\ref{ab}). The form of the function $\phi$
is to be determined from quantum considerations. Since this problem is not yet
solved, in concrete physical applications we have explored several different
possibilities of its form.

The basic variables of the theory under consideration are the spin connection
one-form $A$ and the two-form spinor field $B$, while the Lagrange multiplier
field $\Psi$ can be thought of as expressible in terms of $B$ via the algebraic
``metricity'' equations. The theory with nontrivial $\phi (\beta)$ does not
distinguish any special metric; however, there arises a conformal class of
metrics with respect to which the two-form $B$ is self-dual. The existence of
such a conformal class enables us to speak about the conformal structure of the
solution.

The vacuum spherically symmetric solution has several significant features, and
we can summarize our results as follows:

1.  For arbitrary $\phi$, the theory respects the analog of the Birkhoff
theorem, i.e., the vacuum spherically symmetric solution is necessarily static
and asymptotically describes a space of constat curvature. In our theory, this
means the existence of a vector field generating a group of diffeomorphisms
that leave invariant the fields $A$, $B$ and $\Psi$. In the coordinate
language, the components of these fields do not depend on the ``time''
coordinate $t$.  This property is probably connected with the fact that our
modifications of the Pleba{\'n}ski formulation of the general-relativistic
action does not increase the order of differential equations.  A similar
property was recently proved for metric theories which preserve the
second-order character of field equations \cite{Lovelock}.

2.  In any domain of its argument in which the function $\phi (\beta)$ varies
slowly ($\phi_\beta \ll 1$), it acts simply as a (multiple of) the cosmological
constant, and the solution in the corresponding spatial domain possesses
approximate ``metricity'' property in the sense that the three components of
the self-dual spinor two-form $B$ are expressible as self-dual parts of the
canonical exterior products of some basis one-forms. These basis one-forms then
define a unique metric in the corresponding space-time region, which is an
approximate solution of the vacuum Einstein equations. In our spherically
symmetric solution, such a behavior is obtained at spatial infinity, where we
recover the Schwarzschild metric with small and rapidly decaying ``nonmetric''
corrections. The value of $\beta$ in this case plays the role of Weyl
curvature.

3. If several regions exist in the domain of $\beta$ where the function
$\phi(\beta)$ is slowly varying, then the distinguished metric in the
corresponding spatial regions is described by the Schwarzschild-de~Sitter form
but with different values of the Schwarzschild radius and effective
cosmological constant.  In other words, the gravitational and cosmological
constants become curvature-dependent in this theory. This property of the
solution can potentially be used to account for the problem of missing mass in
spiral galaxies and other astrophysical objects. The nonmetric corrections in
the regions close to the gravitating bodies in this case could also explain the
observed anomaly in the acceleration of the Pioneer spacecraft.

4. In addition to this effective remormalization of gravitational and
cosmological constants, there arises a nontrivial universal redshift factor
between regions of different Weyl curvature $\beta$. This effect potentially
can explain the high redshift of quasars and gamma-ray bursts.

5. The conformal structure of our solution inside the black-hole region is
different from that of the Schwarzschild solution and depends on the form of
the function $\phi$.  Typically, ``inside'' the analog of the Schwarzschild
horizon one finds another surface of extreme ``non-metricity''. This surface is
spacelike, and replaces the usual spacelike singularity inside the
Schwarzschild black hole. The metric ceases to exist at this surface, but all
the dynamical fields of the theory are finite. Thus, this surface is only a
metric singularity, but not a singularity of the theory. Across this surface,
the coordinates $t,r$ change their spacetime roles once more, and one typically
finds another Schwarzschild-like horizon behind this ``non-metricity'' surface.
The theory does not admit arbitrarily small black holes: for small objects, the
curvature on the would-be horizon is so strong that non-metric modifications
prevent the horizon from being formed. Instead of horizon, one has ``naked''
hypersurface of non-metricity in this case.  The details of the conformal
diagram depend on the specific shape of the function $\phi(\beta)$. For a
simple choice $\phi (\beta) = \ell^2 \beta^2$, it is shown in
Fig.~\ref{fig:conf}.

\section*{Acknowledgments}

The authors are grateful to Mart{\'\i}n L{\'o}pez-Corredoira and Jayant
Narlikar for correspondence and to Oleg Barabash for discussion. This work was
done at the Perimeter Institute for Theoretical Physics, whose support the
authors acknowledge. K.~K.\@ was partially supported by an EPSRC advanced
fellowship, and Yu.~S.\@ was supported by grant No.~5-20 of the
``Cosmomicrophysics'' programme and by the Program of Fundamental Research of
the Physics and Astronomy Division of the National Academy of Sciences of
Ukraine, by grant No.~F16-457-2007 of the State Foundation of Fundamental
Research of Ukraine, by the PPARC grant, and by the INTAS grant
No.~05-1000008-7865.

\begin{appendix}

\section{Proof of the static property of the metric}

In this appendix, we give the details of the proof of static property of the
solution to the vacuum spherically symmetric problem.

The most general spherically symmetric expression for the two-form $B$ is given
by (\ref{b}):
\beq\label{Ab}
B = \tlX_- \nm\, m\wedge l + \tlX_+ \nm\, n\wedge \mb + \tlX \left( l\wedge n -
m\wedge \mb \right)
\eeq
with the one-forms $l$, $n$, $m$, and $\mb$ given by (\ref{frame1}):
\beq\label{Aframe1}
l = \frac{1}{\sqrt{2}} \left( fdt - gdr \right)\, , \quad n =
\frac{1}{\sqrt{2}} \left( fdt + gdr \right) \, , \quad m,\, \mb =
\frac{r}{\sqrt{2}} \left( h d\theta \pm i \sin\theta\,  d\phi \right) \, .
\eeq
Now the functions $f$, $g$, $h$, and $\nm$ are not assumed to be
time-independent, but are functions of both $r$ and $t$.  We are going to show
that $h$ is a constant (coordinate-independent), and $f$, $g$, and $\nm$ are
time-independent due to the field equations.

The expressions for $B$ are now modified because of the appearance of time
derivatives, so that instead of (\ref{bs-2}), we have
\beq \label{Abs-2}
\begin{array}{ll}
\displaystyle dB_- = &\displaystyle  {1 \over \sqrt2 r} \left[ {(r \nm f)'
\over g f}\, l \wedge n \wedge m - \nm\, \cot\theta\, l \wedge m \wedge
\mb \right] \medskip \\
&\displaystyle  + {(\nm g)^\cdot \over \sqrt2\, f g}\, l \wedge n \wedge m +
{nm \dot h \over 2 \sqrt2\, f h} l \wedge n \wedge (m + \mb) \, ,
\medskip \\
dB_+ = &\displaystyle  - {1 \over \sqrt2 r} \left[ {(r \nm f)' \over g f}\,
l \wedge n \wedge \mb + \nm\, \cot\theta\, n \wedge m \wedge \mb \right]
\medskip \\
&\displaystyle + {(\nm g)^\cdot \over \sqrt2\, f g}\, l \wedge n \wedge \mb +
{\nm \dot h \over 2 \sqrt2\, f h} l \wedge n \wedge (m + \mb) \, ,
\medskip \\
dB_0 = &\displaystyle \frac{\sqrt2}{r g}\, (l - n) \wedge m \wedge \mb - {\dot
h \over \sqrt2\, f h}\, (l + n) \wedge m \wedge \mb \, ,
\end{array}
\eeq
where overdot denotes the time derivative.  Formulas (\ref{db-comp}) are then
modified as follows:
\beq \label{Adb-comp}
\begin{array}{l}
\displaystyle \left( dB_- \wedge m \right)_v = - {\nm \dot h \over 2 \sqrt2\, f
h} \, , \quad \left( dB_- \wedge \mb \right)_v = \frac{(r \nm f)'}{\sqrt{2}\, r
g f} + { (\nm g)^\cdot \over \sqrt2\, f g} + {\nm \dot h \over 2 \sqrt2\, f h}
\, , \medskip \\
\displaystyle \left( dB_+ \wedge m \right)_v = \frac{(r \nm f)'}{\sqrt{2}\, r g
f} - { (\nm g)^\cdot \over \sqrt2\, f g} - {\nm \dot h \over 2 \sqrt2\, f h} \,
, \quad \left( dB_+ \wedge \mb \right)_v = {\nm \dot h \over 2 \sqrt2\, f h} \,
, \medskip \\
\displaystyle - \left( dB_- \wedge n \right)_v = \left( dB_+ \wedge l \right)_v
= \frac{\nm}{\sqrt2\, r} \cot\theta \, ,
\medskip \\
\displaystyle \left( dB_0 \wedge n \right)_v = \frac{\sqrt2}{r g} - {\dot h
\over \sqrt2\, f h} \, , \quad \left( dB_0 \wedge l \right)_v = \frac{\sqrt2}{r
g} + {\dot h \over \sqrt2\, f h} \, ,
\end{array}
\eeq the rest of the projections being zero.  Then, using formulas
(\ref{c-comp}) and (\ref{a-comp}), which remain unchanged, we obtain
\beq\label{Aa}
\begin{array}{c}
\displaystyle A_- = \nm^{-1} P m + \nm Q \mb \, , \qquad
A_+ = \nm Q m + \nm^{-1} P \mb  \, , \medskip \\
\displaystyle A_0 = \left( \nm^{-2} P + R \right) (l + n) - {\cot\theta \over
\sqrt2\, r h}\, (m - \mb) + {\dot h \over \sqrt2\, \nm^2 f h} \, l - { \left(
\nm g \sqrt{h} \right)^\cdot \over \sqrt2\, \nm f g \sqrt{h}}\, (l - n)  \, ,
\end{array}
\eeq
where we made the notation
\beq \label{pqr}
\begin{array}{l}
\displaystyle P = - {1 \over 2 \sqrt2\, g} {\left(r^2 h \right)' \over r^2 h} -
{\dot h \over 2 \sqrt2\, fh} \, , \medskip \\
\displaystyle Q = - {h' \over 2 \sqrt2\, g h} - {\dot h \over 2 \sqrt2\, f h}
\, , \medskip \\
\displaystyle R = {\left(r \nm f \sqrt{h}\right)' \over \sqrt2\, r \nm f g
\sqrt{h}} \, .
\end{array}
\eeq

Now we have to compute the left-hand sides of equations (\ref{eqs-1}) and
equate them to zero.  Computing the $m \wedge \mb$ component of any of the
first two equations in (\ref{eqs-1}), we immediately obtain that $Q = 0$,
which, in turn, simplifies the expressions (\ref{Aa}).  Calculating then the
$(l \wedge \mb)$ and $n \wedge \mb$ components of the first equation in
(\ref{eqs-1}), we get
\beq
P \left({h' \over gh } - {\dot h \over fh} \right) = 0 \, , \qquad P \left({h'
\over gh } + {\dot h \over fh} \right) = 0 \, ,
\eeq
respectively. The condition $P = 0$ is excluded because it contradicts the $l
\wedge m$ component of the same equation, which contains a generically nonzero
expression $\beta + \phi$.  Therefore, one must have $h \equiv {\rm const}$.

The remaining four equations stemming from the first two equations in
(\ref{eqs-1}) read
\beq \label{four}
\begin{array}{l}
\displaystyle {\left( r \nm^{-1} P \right)' \over \sqrt2\, rg }  - { P \dot g_*
\over  \sqrt2\, \nm f g_* } + \nm^{-1} P \left( \nm^{-2} P + R \right) - {
\left( \nm^{-1} P \right)^\cdot \over \sqrt2\, f} = - \nm ( \beta + \phi)
\, , \medskip \\
\displaystyle {\left( r \nm^{-1} P \right)' \over \sqrt2\, rg }  - { P \dot g_*
\over  \sqrt2\, \nm f g_* } - \nm^{-1} P \left( \nm^{-2} P + R \right) + {
\left(
\nm^{-1} P \right)^\cdot \over \sqrt2\, f} = 0 \, , \medskip \\
\displaystyle {\left( r \nm^{-1} P \right)' \over \sqrt2\, rg }  + { P \dot g_*
\over  \sqrt2\, \nm f g_* } + \nm^{-1} P \left( \nm^{-2} P + R \right) + {
\left(
\nm^{-1} P \right)^\cdot \over \sqrt2\, f} = - \nm ( \beta + \phi) \, , \medskip \\
\displaystyle {\left( r \nm^{-1} P \right)' \over \sqrt2\, rg }  + { P \dot g_*
\over  \sqrt2\, \nm f g_* } - \nm^{-1} P \left( \nm^{-2} P + R \right) - {
\left( \nm^{-1} P \right)^\cdot \over \sqrt2\, f} = 0 \, ,
\end{array}
\eeq
where $g_* = \nm g$, as usual.  Now, subtracting the first equation from the
third one, and the second from the fourth one, we obtain, respectively,
\beq
{P \dot g_* \over \nm g_*} + \left( \nm^{-1} P \right)^\cdot = 0 \, , \qquad {P
\dot g_* \over \nm g_*} - \left( \nm^{-1} P \right)^\cdot = 0 \, ,
\eeq
which implies
\beq
\label{static} \dot g_* = 0 \, , \qquad \left( \nm^{-1} P \right)^\cdot = 0
\, .
\eeq
These two equations are equivalent in view of the condition $h = {\rm const}$
and definition (\ref{pqr}).

Under condition (\ref{static}), the remaining two equations stemming from
(\ref{four}) are precisely equations (\ref{first}), while the last equation in
(\ref{eqs-1}) leads to the two equations (\ref{second}).  Thus, we have the
system of differential equations (\ref{system}).  The second equation in
(\ref{system}) then implies that $\beta$ does not depend on time, hence, by
virtue of (\ref{c}), $\nm$ is also time-independent.  Finally, the third
equation in (\ref{system}) implies that $f$ can only have a time-dependent
overall factor, which can always be rescaled to a constant by changing the time
variable.  This completes the proof of the static property of the spherically
symmetric vacuum solution in the theory under investigation.

\section{Mass renormalization and redshift: general analysis}
\label{app:general}

In this appendix, we generalize the examples considered in Sec.~\ref{sec:mod}.
Thus, assume that two regions are characterized by the conditions $\ell^2 \beta
\ll 1$ and $\ell^2 \beta \gg 1$, respectively, where $\ell$ is some length
scale.  We will find the approximate solutions in these regions assuming the
behavior
\beq
\phi_\beta = \left\{
\begin{array}{rl}
2 \alpha_1 x^{p_1} + o \left( x^{p_1} \right) \, , \quad &x \ll 1 \, ,  \\
2 \alpha_2 x^{-p_2} + o \left( x^{-p_2} \right) \, , \quad &x \gg 1 \, ,
\end{array}
\right.
\eeq
where the variable $x = \ell^2 \beta$, and $\alpha_1$, $\alpha_2$, $p_1 > 0$
and $p_2 > 0$ are different constants.

At small radial distances, where $x \gg 1$, we integrate equations (\ref{new})
to obtain
\ber
{\ell^2 r_s \over 2 r^3} = x \left[ 1 + {\alpha_2 \over p_2} x^{-p_2}  + o
\left(x^{-p_2} \right) \right] \, , \quad f_* g_* = 1 + {\alpha_2
\over p_2} x^{-p_2}  + o \left(x^{-p_2} \right) \, , \nonumber \\
x \gg 1 \, .
\eer
We also have
\beq
\nm^2 = 1 - 3 \alpha_2 x^{-p_2} + o \left(x^{-p_2} \right) \, , \quad x \gg 1
\, .
\eeq
Collecting all the terms, we obtain the leading contribution to the $g_{00}$
coefficient in metric (\ref{metric}):
\beq \label{small-gen}
g_{00} = f^2 = 1 - {r_s \over r} + \phi_\infty r^2 + 3 \alpha_2 \left(1 + {2
\over 3 p_2} \right) \left( { 2 r^3 \over \ell^2 r_s} \right)^{p_2} \, , \qquad
r^3 \ll \ell^2 r_s \, ,
\eeq
where $r_s = \mbox{const}$ is the value of the effective Schwarzschild radius
at small distances, $\phi_\infty$ is the value of the function $\phi(\beta)$ at
infinity. We have also assumed the condition $r_s \ll r$, which guarantees
that some terms that are of lower power in $r$ are actually subleading. This expression
is valid if $p_2 \ne 1$. In the interesting case $p_2 = 1$, the correction to
the Schwarzschild--de~Sitter metric will also have logarithmic terms; see our
concrete example (\ref{phi-log}) in Sec.~\ref{sec:mod}.

As we discussed above, the physical metric can be specified by an additional
conformal factor, in which case, we obtain somewhat different corrections.
Thus, for the Urbantke metric defined in (\ref{urbantke}) and for the
``volume'' metric defined in (\ref{metric-V}), we will have the coinciding
approximate relations for $r^3 \ll \ell^2 r_s\,$:
\beq \label{small-alt}
\tilde g_{00} = \displaystyle 1 - {r_s \over r} + \phi r^2 + 2 \alpha_2 \left(1
+ {1 \over p_2} \right) \left( { 2 r^3 \over \ell^2 r_s} \right)^{p_2}  \, .
\eeq

At large radial distances, where $x \ll 1$, we have
\ber
{\ell^2 r_s \over 2 r^3} = \exp \left( \int_0^\infty {\phi_\beta \over 2 \beta}
d \beta \right) x \left[ 1 - {\alpha_1 \over p_1} x^{p_1}  + o \left(x^{p_1}
\right) \right] \, , \nonumber \\
f_* g_* = \exp \left( \int_0^\infty {\phi_\beta \over 2 \beta} d \beta \right)
\left[ 1 - {\alpha_1 \over p_1} x^{p_1}  + o \left(x^{p_1} \right) \right] \, ,
\eer
\beq
\nm^2 = 1 - 3 \alpha_1 x^{p_1} + o \left(x^{p_1} \right) \, , \quad x \ll 1 \,
.
\eeq
In this case, we obtain the following leading contribution to the metric:
\beq \label{large-gen}
g_{00} = f^2 = Z^{-2} \left[ 1 - {Z r_s \over r} + \phi_0 r^2 + 3 \alpha_1
\left(1 - {2 \over 3 p_1} \right)  \left( {\ell^2 Z r_s \over 2 r^3 }
\right)^{p_1} \right] \, , \qquad r^3 \gg \ell^2 Z r_s \, ,
\eeq
where $\phi_0 = \phi(0)$, and
\beq \label{z-gen}
Z = Z(\infty, 0) = \exp \left( - \int_0^\infty {\phi_\beta \over 2 \beta} d
\beta \right) \, .
\eeq

In the case of extra conformal factors, again, the corrections to the
Schwarzschild-de~Sitter metric will be somewhat different. For the Urbantke
metric defined in (\ref{urbantke}) and for the ``volume'' metric defined in
(\ref{metric-V}), we have the same approximate relations for $r^3 \gg \ell^2
r_s\,$:
\beq \label{large-alt}
\tilde g_{00} = Z^{-2} \left[ 1 - {Z r_s \over r} + \phi_0 r^2 + 2 \alpha_1
\left(1 - {1 \over p_1} \right)  \left( {\ell^2 Z r_s \over 2 r^3 }
\right)^{p_1}  \right] \, .
\eeq

We note that $\nm^2 \approx 1$ in our approximation in these regions, so the
physical metric is well defined. The contributions proportional to $\alpha_2$
and $\alpha_1$ in (\ref{small-gen}), (\ref{small-alt}) and (\ref{large-gen}),
(\ref{large-alt}), respectively, are small and can be dropped if the constants
$\alpha_1$ and $\alpha_2$ are not very large. However, these contributions
themselves are of the same order as the ``nonmetricity.'' Their physical
interpretation, therefore,  requires the knowledge of the matter couplings in
our theory, which is an issue still to be resolved.

For $Z > 1$, the observed gravitational mass of the central object at large
distances is $Z$ times larger than it is at small distances.  In the
intermediate region of radial distances
\beq \label{between-gen}
1 \lesssim {r^3 \over \ell^2 r_s } \lesssim Z \, ,
\eeq
the metric coefficient behaves in a complicated way; moreover, the physical
metric may not be well defined in this region at all.  If several regions exist
in the space of $\beta$ in which $\phi_\beta \ll 1$,
then obvious renormalizations of the observed gravitational masses and redshift
factors exist between these regions.

\end{appendix}

\end{document}